\documentclass{aa}
\usepackage{natbib}
\usepackage{txfonts}
\usepackage{graphicx}  
\bibpunct{(}{)}{;}{a}{}{,}
\usepackage[english]{babel}
\usepackage{ulem}
\usepackage{array}
\usepackage{hyperref}
\usepackage{psfrag}

\newcommand{\Porb}{\mbox{$P_\mathrm{orb}$}}

\newcommand{\Msun}{\mbox{$\mathrm{M_{\odot}}$}}

\newcommand{\Teff}{\mbox{$T_\mathrm{eff}$}}

\newcommand{\Mtr}{\mbox{$\dot{M}_{\mathrm{tr}}$}}
\newcommand{\Mdot}{\mbox{$\dot{M}$}}

\newcommand{\Mcrp}{\mbox{$\dot{M}^+_{\mathrm{cr}}$}}
\newcommand{\Mcrm}{\mbox{$\dot{M}^-_{\mathrm{cr}}$}}

\def\apgt{\ {\raise-.5ex\hbox{$\buildrel>\over\sim$}}\ }
\def\aplt{\ {\raise-.5ex\hbox{$\buildrel<\over\sim$}}\ }

\newcommand{\ah}{\mbox{$\alpha_{\mathrm{h}}$}}
\newcommand{\ac}{\mbox{$\alpha_{\mathrm{c}}$}}

\newcommand{\ptf}{PTF~1J0719+4858\,}

\newcommand {\be} {\begin{equation}}
\newcommand {\ee} {\end{equation}}
\newcommand{\beqa}{\begin{eqnarray}}
\newcommand{\eqa}{\end{eqnarray}}
\newcommand{\bea}{\begin{eqnarray}}
\newcommand{\eea}{\end{eqnarray}}
\newcommand {\bc} {\begin{center}}
\newcommand {\ec} {\end{center}}
\newcommand{\nn}{\nonumber}

\def\spose#1{\hbox to 0pt{#1\hss}}
\def\simless{\mathrel{\spose{\lower 3pt\hbox{$\mathchar"218$}}
        \raise 2.0pt\hbox{$\mathchar"13C$}}}
\def\simgreat{\mathrel{\spose{\lower 3pt\hbox{$\mathchar"218$}}
        \raise 2.0pt\hbox{$\mathchar"13E$}}}
\def\lta{\mathrel{\spose{\lower 3pt\hbox{$\mathchar"218$}}
        \raise 2.0pt\hbox{$\mathchar"13C$}}}
\def\gta{\mathrel{\spose{\lower 3pt\hbox{$\mathchar"218$}}
        \raise 2.0pt\hbox{$\mathchar"13E$}}}

\begin{document}

\title{Models of AM CVn star outbursts}

\author{Iwona Kotko\inst{1}
\and
              Jean-Pierre Lasota\inst{1,2}
\and       Guillaume Dubus\inst{3,2}
\and       Jean-Marie Hameury\inst{4}
}
\offprints{I.Kotko@oa.uj.edu.pl}

\institute{Astronomical Observatory, Jagiellonian University, ul. Orla 171, 30-244 Krak\'ow, Poland
             \and
               Institut d'Astrophysique de Paris, UMR 7095 CNRS, UPMC Univ Paris 06, 98bis Bd Arago, 75014 Paris, France
                \and
                 UJF-Grenoble1/\,CNRS-INSU, Institut de Plan\'etologie et d'Astrophysique de Grenoble (IPAG) UMR 5274, Grenoble, F-38041, France
                 \and
                 Observatoire de Strasbourg, CNRS/Universit\'e de Strasbourg, 11 rue de l'Universit\'e, F-67000 Strasbourg, France
              }
\abstract{Outbursting AM~CVn stars exhibit outbursts similar to those observed in different types of dwarf novae. Their light-curves combine the characteristic features of SU~UMa, ER~UMa, Z~Cam, and WZ~Sge-type systems but also show a variety of properties never observed in dwarf novae. The compactness of AM~CVn orbits and their unusual chemical composition make these systems valuable testbeds for outburst models.}
{We aim for a better understanding of the role of helium in the accretion disc instability mechanism, testing the model for dwarf novae outbursts in the case of AM~CVn stars, and aim to explain the outburst light-curves of these ultra-compact binaries.}
{We calculated the properties of the hydrogen-free AM~CVn stars using our previously developed numerical code adapted to the different chemical composition of these systems and supplemented with formulae accounting for mass transfer rate variations, additional sources of the disc heating, and the primary's magnetic field.}
{We discovered how helium-dominated discs react to the thermal-viscous instability and were able to reproduce various features of the outburst cycles in the light-curves of AM~CVn stars.}
{The AM~CVn outbursts can be explained by the suitably adapted dwarf-nova disc instability model but, as in the case of its application to hydrogen-dominated cataclysmic variables, one has to resort to additional mechanisms to account for the observed superoutbursts, dips, cycling states, and standstills. We show that 
the enhanced mass-transfer rate, due presumably to variable irradiation of the secondary, must not only be taken into account but is a determining factor that shapes AM~CVn star outbursts. The cause of the variable secondary's irradiation has yet to be understood; the best candidate is the precession of a tilted/warped disc.}
\maketitle

\section{Introduction}
\label{sect:intro}
AM~CVn stars (AM~CVns) are binary systems with very short orbital period ($10-65\,\,{\rm min}$) 
in which a He or C/O white dwarf primary accretes matter lost by a Roche-lobe filling
secondary star, which is believed to be also a helium white dwarf \footnote{The nature of the two shortest orbital period,
(X-ray pulsating) systems HM Cnc and V407 Vul remains a mystery.}\citep{Nel05}.
The observed optical spectra show that accretion discs in
these systems are helium-dominated and hydrogen-free. This spectral feature is indeed the criterion for including an ultra-compact binary to the  AM~CVn class. AM~CVns are observed
in three distinct luminosity states \citep[see e.g.][]{Bild06}.
Two of these are the persistent--luminous (high) and the outbursting states, which are similar to those seen in 
cataclysmic variable stars (CVs) in which
the secondary is hydrogen-rich and usually non-degenerate. The third state, during which the system is persistent
and faint, does not have an equivalent among CVs. \citet{Smak83} remarked that the disc instability model
\citep[DIM; see][for a review and references]{L01} which describes dwarf novae (DN) also applies 
to AM~CVn stars and  can explain the properties of AM~CVn itself (high state) and of GP~Com (faint state). Because the
instability in question is triggered by opacity changes, the differences between AM~CVns and
CVs are supposed to be explained mainly by the higher ionization potential of helium compared to that of hydrogen
\citep[see][]{can84,TO97,LDK}. However, as we show in the present paper, other effects related to the helium
atomic structure as well as the presence of metals may also influence the outburst cycle
properties. 

In this context it is also  worth stressing that angular momentum transport in discs depends on the microphysics and may be significantly affected by the chemical composition (e.g. the chemical composition influences the Prandtl number, which affects the development of the magneto-rotational instability). One of the methods to test the disc angular momentum transport mechanisms is outburst modelling.

Although still poor compared to that of CVs, the AM~CVns observation database has been
enriched by recent observations and we know more about these binaries than we did in 1997. Then, in their
pioneering work, \citet{TO97} used a highly simplified description of the DIM \citep[][]{O89} and not the full  
DIM model \citep[such as e.g.][]{IO92}.

These facts as well as the desire to apply our version of the DIM
for helium-rich accretion discs \citep[][]{LDK,KLD1} to AM~CVn outbursts has been the main motivation for
writing the present article. 
In Section \ref{sect:Outb.AMCVn1} we discuss the general properties of AM~CVn outbursts,
concentrating on those we are trying to explain in the following parts of the article. 
Section \ref{sect:DIMapp} describes
the DIM in the context of helium-dominated accretion discs. 
Section \ref{sect:mtranstab} briefly recalls what is known about the chemical composition of AM~CVn discs and how they comply with the stability criteria set by the DIM. 
In Section \ref{sect:models} we investigate the properties of helium-rich disc models in general and in Sections \ref{sect:modellc} and  \ref{sect:super} we apply them to various types of AM~CVn star outbursts. We discuss our results and summarize the conclusions in \ref{sect:concl}.

\section{Outburst properties of AM~CVn stars}
\label{sect:Outb.AMCVn1}

From among 27 systems classified as AM~CVn binaries, 11 have been observed to produce outbursts \citep[][]{Levitan11a,Rams11}. Because of their relative 
faintness ($\sim 14 - 20$ mag) the light-curves of these systems are not as well covered and sampled as those of the longer period, 
hydrogen-dominated CVs. Based on a very incomplete (but soon to be increased by observational campaigns such as the Palomar Transient Factory, \citet{Levitan11b}) AM~CVn database one can conclude that most 
(all ?) outbursting helium-rich systems show superoutbursts or, to be more precise, outbursts that resemble superoutbursts observed 
in hydrogen-rich dwarf novae such as SU UMa stars \citep[][]{Warner03}. The most important similarity is the presence of 
superhumps - photometric humps  with periods slightly different from that of the orbital modulation. The amplitudes of these outbursts range 
from 3.5 to 6 mag; the recurrence times from $\sim 45$ to $\sim 450$ days, but this latter parameter has been determined only for five
systems with orbital periods between 24.52 and 28.32 min. 
AM~CVn with longer orbital periods are believed \citep[e.g.][]{Levitan11a} to have recurrence times longer 
than three months but this still  requires confirmation. One should also stress that in two cases a substantial variation of the recurrence time has 
been observed: in CR~Boo from 46.3 to 14.7 days \citep[][]{Kato2001} and in KL~Dra from $\sim $ 65 to $\sim $ 44 days \citep[][]{Rams10,Rams11}. 
The superoutbursts of AM~CVn stars typically last about 20 days but can be as short as 9 days. On the other hand, in some cases \citep[see e.g.][]{Patt1997,Patt2000}
superoutbursts are prolonged by a series of low-amplitude, frequent outbursts: the so-called {\sl cycling state} that may last as 
long as the ``smooth" superoutburst itself. Often the superoutburst is in fact not so smooth because it contains a fairly pronounced dip 
\citep[][]{Rams10,Rams11}. The relation, if any, of these dips to the cycling state and of these two features to the ``dipping state" 
observed in the light-curves of some of  WZ~Sge-type stars has not been satisfactorily elucidated yet.  

Much less in known about the so-called ``normal outbursts" that are common in hydrogen-rich dwarf novae, but rarely observed in 
outbursting AM~CVn stars.

From the observational point of view, ``normal" outbursts are not very well defined but they are supposed to correspond to the quite narrow eruptions observed in U~Gem-type dwarf novae. They are  defined in opposition to the superoutbursts whose most characteristic feature are superhumps. Normal outbursts could therefore be defined by the lack of superhumps (but in some cases normal outbursts following a superoutburst do show superhumps.) 

In the case of CVs, no short period (with orbital periods below the 2 -- 3 hr ``period gap") dwarf nova shows a pure normal outburst cycle. The SU UMa stars show a supercycle in which normal outbursts appear between superoutbursts, and binaries of the WZ~Sge-type show superoutbursts only. The shortness of period seems to be the decisive factor determining the presence of superoutbursts\footnote{How short is the required period is subject to debate since U~Gem, at about 4 hr orbital period, showed a superoutburst \citep[][]{Masonetal88,SmakWaagen04,Smak06,Schreib07}} which is confirmed by the fact that all AM~CVns exhibit these types of outbursts.

Noticing that in CR~Boo and V803~Cen  the outbursts seen during the cycling state satisfy the empirical Kukarkin-Parenago 
relation, connecting the amplitude $A_{n}$ with the recurrence time $T_{n}$ \citep[see][and Sect. \ref{KP}]{KL1}, \citet{Patt1997,Patt2000} concluded that they can be qualified as normal outbursts. On the other hand, \citet[][]{Kato2001} considered this state as equivalent to the standstill in Z~Cam type dwarf novae. Until now
the only clear case for relatively well sampled normal outbursts has been provided by observations of  PTF1~J0719+4858 \citep{Levitan11a}. In this
case the amplitudes are $\sim 2.5$ mag, the duration $\sim 1$ day and the recurrence time about 10 days. In this system normal outbursts 
appear between superoutbursts as in SU~UMa stars. \citet{Levitan11a} suggested that AM~CVn with orbital periods shorter than $\sim 27$~min are
equivalent to SU~UMa stars while helium-rich systems with longer periods would correspond to long recurrence time WZ~Sge stars showing only 
superoutbursts. Although not implausible, this hypothesis has to await a more complete observational coverage of AM~CVn binaries. Let us
just mention that a complete analogy between dwarf novae and their helium-rich cousins might be difficult to establish since e.g. the short-period KL Dra 
seems to exhibit only (short-recurrence period) superoutbursts \citep[see, however,][]{Rams11}.

\subsection{The DIM for  AM~CVn outbursts: preliminary considerations}
\label{sect:prelim}

A model accounting for the properties of the three categories of AM~CVn stars - permanently bright, permanently faint, and outbursting has to take into 
account the two main differences between these systems and the hydrogen-dominated CVs: the helium-dominated accretion discs and shorter orbital 
periods, that is, shorter (less extended) accretion discs. As noticed by \citet{Smak83} the obvious candidate here is the DIM that is used to explain 
outbursts of hydrogen-rich dwarf novae. It might seem that the adaptation of this model to the case of AM~CVn is pretty straightforward because it should 
consist only in rescaling to helium-rich discs the critical values defining the instability strip. This simplistic view assumes, however, that the DIM is  fully 
successful in explaining most of the dwarf nova outburst properties. Unfortunately this is not the case. 

First, the standard DIM does not account for superoutbursts, which is particularly embarrassing for AM~CVn stars in which this type of 
outburst is observed most frequently.  Two modifications of the DIM have been proposed to account for superoutbursts of dwarf novae.  \citet
{O89} \citep[see also][for a review]{O96} proposed that superoutbursts in SU UMa stars are produced by a tidal-thermal instability (TTI) triggered by 
a large enhancement of the tidal torque acting on the disc. Despite successes in describing some of the superoutburst properties, the TTI model has 
encountered some insurmountable difficulties  when trying to agree with observations \citep[see e.g.][and references therein]{BH02,SHL04,HL05}. 
Recently, in a remarkable series of articles \citet{Smak09a,Smak09b,Smak09c,Smak09d} has undermined the basis of the TTI model by casting doubt 
on the presence of an eccentric disc in the SU~UMa stars and providing strong observational arguments in favour of a different model: the enhanced 
mass-transfer (EMT) model, originally proposed by \citet{O85} and developed by \citet{LHH95}, \citet{BH02}, \citet{SHL04} and \citet{HL05}.  According to the EMT model: 
\textsl{superoutbursts are due and begin with a major enhancement in the mass transfer rate. During the "flat-top" part of the superoutburst the mass 
transfer rate decreases slowly, causing the observed luminosity to decline. The superoutburst ends when the mass transfer rate decreases below its 
critical value, resulting in a transition to the quiescent state of the dwarf nova cycle} \citep{Smak08}.
In this article we will be testing the EMT superoutburst model for several reasons.

First, there is (mostly indirect) evidence of large variations of the mass-transfer rate at least in some outbursting AM~CVn \citep[see e.g.][]{Patt2000}. Second, as shown by \citet[][]{SHL04}, the differences between standard SU~UMa stars and ER~UMa systems are a natural outcome of the EMT, but not of the TTI model therefore it is unlikely that the latter will be able to describe a system behaving like a WZ~Sge, SU~UMa, ER~UMa, and Z~Cam star. Third, since (unchallenged) \citet[][]{Smak09a} has undermined the very basis of the TTI model for the SU~UMa stars it is normal to assume it is also not relevant to AM~CVn star outbursts.
 
Although normal outbursts seem to be infrequent in AM~CVn stars, they are the phenomena the DIM is supposed to describe and explain. Since the 
standard DIM can be tested only on such outbursts, there is an obvious difficulty in applying this model to AM~CVns. The recent observation of AM~CVn star outbursts that can be clearly identified as normal \citep[][]{Levitan11a} promises a more straightforward comparison with the DIM predictions. In particular, it 
might help solving one important problem with this model: the scaling and re-scaling of the viscosity parameter $\alpha$. As first noted by \citet
{Smak84}, to obtain dwarf nova outbursts with observed amplitude the viscosity parameter during outburst $\alpha_{\mathrm{h}}$ must be 
at least four times larger than the quiescent $\alpha_{\mathrm{c}}$. When implemented, this ad hoc assumption increases the ratio of critical surface densities which un turn 
allows (because of the S-shaped equilibrium curves) adequately large outburst amplitudes. 

As showed by \citet{Smak99} for dwarf novae, $\alpha_{\mathrm{h}}\approx 0.2$. He arrived at this conclusion in two ways. First, by comparing the observed relation between the normal-outburst decay time and the orbital period ($t_{\rm dec}-P_{\rm orb}$) with the model-calculated decay time $t_{\rm dec}= R_D/v_R$, 
where $R_D$ is the outer disc radius and $v_R$ the viscous speed in the hot disc. Second,  by using the normal-outburst width vs orbital period ($W - P_{\rm orb}$) relation to compare with the width of the normal outbursts obtained from the model. The viscous speed is
\begin{equation}
v_{R}\sim \frac{\nu}{R}\sim \alpha_{\mathrm{h}}c_s^2 v_K^{-1}\sim \alpha_{\mathrm{h}}\frac{\gamma kT_c}{\mu m_H} R^{1/2},
\label{eq:vvist}
\end{equation}
where $\nu$ is the kinematic viscosity coefficient, $c_s$ the sound speed, $v_K$ the Keplerian velocity, $\mu$ the mean molecular weight, $\gamma$ the adiabatic index, $m_H$ the hydrogen atomic mass, and $T_c$ is the mid-plane temperature.
Therefore the viscous speed in AM~CVn discs is half that of hydrogen-dominated CVs (the critical temperatures, for which ionization/recombination of the dominant element in the disc takes place, are twice as high
but the molecular weight is four times higher). Since AM~CVn discs are shorter (by a factor $\sim 2$) than those of  CVs, decay times in the former should be comparable to those in the latter. 
The available limited sample seems to confirm that.  
The correlation found by \citet{Smak99} between the observed outburst widths $W(d)$ (in days) 
and the orbital period $\Porb$ (in hours) is $W(d)=(2.01\pm 0.29)\Porb(\rm hr)^{(0.78\pm 0.11)}$. Therefore for $\Porb = 25$~min the normal outburst duration should be $\sim 1$, day as observed in 
PTF~1J0719+4858 by \citet{Levitan11a}. Based on this very limited evidence, one can conclude that the value of the viscosity parameter in hot AM~CV discs is close to that of hydrogen-rich dwarf novae \citep{KL1}. For convenience we will use the two fiducial values $\alpha_{\mathrm{h}} = 0.1$ and $\alpha_{\mathrm{h}} = 0.2$, hoping that observations will soon bring more information. The ratio $\ah/\ac$ required to reproduce observed properties of AM~CVn outbursts have to be determined through the DIM calculations. This problem will be addressed below. First, we have to establish how the change of chemical composition affects the DIM.

\section{The disc instability model for AM~CVn stars: application}
\label{sect:DIMapp}

The model assumes that the disc is geometrically thin, which allows decoupling
the radial and vertical structure equations. The radial structure
of the disc is described by the equations of mass, angular momentum
and energy conservation \citep[see Eqs. (1), (2), (4) in][hereafter H98]{HMDLH} and its vertical
structure is assumed to be in hydrostatic equilibrium, which is described
by equations (14-16) and (23) from H98.

By solving the equations of the disc local vertical structure one
obtains the values of effective temperature ${T}_{\rm eff}$ (or central
temperature $T_{c}$) and surface density $\Sigma$, for which the
thermal equilibrium, at a given disc radius ${R}$, is calculated.
At each ${R}$ these solutions form on the $T-\Sigma$ plane the well-known S-shaped curve.
The lower branch of this curve, terminating at a maximum value of the surface density $\Sigma_{\rm crit}^{-}$,
represents stable cold disc equilibria, while the upper branch starting at a critical
minimum value of the surface density $\Sigma_{\rm crit}^{+}$, corresponds to stable, hot equilibrium
disc solutions. Solutions on the middle branch are thermally unstable. Their existence is the cause of the dwarf~nova outbursts.

For a pure helium disc ($Y=1$) the critical surface densities $(\Sigma_{\rm crit}^{\pm})$ and the critical
effective temperatures $(T_{\rm eff}^{\pm})$ can be fitted by the formulae found by Lasota et al.
(2008):
\beqa
\Sigma_{\rm crit}^{+} & =& 528~\alpha_{0.1}^{-0.81}~R_{10}^{ 1.07}~M_1^{-0.36}\, \rm g\,cm^{-2}\\
T_{\rm eff}^{+} & =& 13000~\alpha_{0.1}^{-0.01}~R_{10}^{-0.08}~M_1^{ 0.03}\,\rm K \\
\Sigma_{\rm crit}^{-} & =& 1620~\alpha_{0.1}^{-0.84}~R_{10}^{ 1.19}~M_1^{-0.40}\,\rm g\,cm^{-2}\\
T_{\rm eff}^{-} & =& 9700~\alpha_{0.1}^{-0.01}~R_{10}^{-0.09}~M_1^{ 0.03}\,\rm K ,\\
\nn
\label{eq:he_nonirr_crit}
\eqa
whereas for a helium disc with solar metal abundance ($Z=0.02$, $Y=0.98$) one has
\beqa
\Sigma_{\rm crit}^{+} & =& 380~\alpha_{0.1}^{-0.78}~R_{10}^{ 1.06}~M_1^{-0.35}\, \rm g\,cm^{-2}\\
T_{\rm eff}^{+} & =& 11500~\alpha_{0.1}^{-0.01}~R_{10}^{-0.08}~M_1^{ 0.03}\,\rm K \\
\Sigma_{\rm crit}^{-} & =& 612~\alpha_{0.1}^{-0.82}~R_{10}^{ 1.10}~M_1^{-0.37}\,\rm g\,cm^{-2}\\
T_{\rm eff}^{-} & =& 8690~\alpha_{0.1}^{-0.00}~R_{10}^{-0.09}~M_1^{ 0.03}\,\rm K, \\
\nn
\label{eq:he_nonirr98_crit}
\eqa
where $\alpha_{0.1}$ is the viscosity parameter in units of $0.1$,
$R_{10}$ is the disc radius in $10^{10}\,\,{\rm cm}$ and $M_{1}$ the white dwarf primary's
mass in solar units (see Appendix \ref{app1} for the complete list of critical values, also for $Z=0.04$).

The critical effective temperatures are independent of the viscosity parameter $\alpha$ 
(the very weak dependence in some cases results from the imprecision of the fit) because in thermal
equilibrium the flux (hence $T_{\rm eff}$) is independent of the viscosity mechanism \citep[see e.g.][]{fkr}.
This allows a fairly simple understanding of the S-curve properties.

From mass and angular-momentum conservation equations one obtains
\begin{equation}
\label{eq: TE }
\nu \Sigma \sim \sigma T_{\rm eff}^4.
\end{equation}
The kinematic viscosity coefficient is $\nu=2/3\,\alpha\,c_s\,H$, where $c_s$ is the speed of sound and $H$ the disc semi-scale-height.
Therefore (for a gas-pressure-dominated disc) one has
\begin{equation}
\label{eq:tcsigma}
\alpha T_c \Sigma \sim \sigma T_{\rm eff}^4 ,
\end{equation}
which implies that for $T_c\propto \alpha^n$ and $\Sigma\propto \alpha^m$, $n + m = -1$. Eq.~({\ref{eq:tcsigma}) and 
the energy transfer equation linking $ T_c$ and $T_{\rm eff}$, determine the slopes of the S-curve's branches $\Teff(\Sigma)$.

In the case of radiative cooling the effective and central temperatures are related through
\be
T_{\rm c}={\left(\frac{3\tau_{\rm tot}}{8}\right)^{1/4}} \Teff
\label{eq:radeq}
\ee
where $\tau_{\rm tot}$ is the total disc optical depth \citep[see][]{dubetal-99}. 
For a fully ionized disc the Rosseland opacity coefficient can be written as $\kappa \sim \Sigma H^{-1} T_c^{-7/2}
\,\mathrm{cm^2/g}$ from which follows the well-known relation $\Teff \propto \Sigma^{5/14}$ found in the
classical \citet{SS73} solution. On the other
hand, a very steep decrease of opacity with temperature induces a change of slope of the $\Teff(\Sigma)$ relation.
This is what happens during recombination.

Fig. \ref{fig:scurve} shows S-curves of discs with four chemical compositions: $Y=1.0, Y=0.98, Y=0.96$, and
solar abundance ($X=0.7, Y=0.28, Z=0.02$). As expected, for all cases the upper branches  are parallel (and very close). 
Since the change of slope at $\Sigma_{\rm crit}^{+}$ 
is caused by a recombination-induced change in the opacity dependence on temperature, it follows that the upper S-curve bend will be the highest
for the element with the highest ionization potential.
Hence $\Sigma_{\rm crit}^{+}(Y=1)> \Sigma_{\rm crit}^{+}(Y=0.98)>\Sigma_{\rm crit}^{+}(Y=0.96)> \Sigma_{\rm crit}^{+}(\mathrm{solar})$.
Cooling by convection, which flattens the temperature profile, affects the position of the upper bend of the S-curve, but not the relative
order of the critical surface-densities values.

For the lower bend of the S-curve, the situation is slightly more complicated. According to \citet{TO97} it
appears where the disc becomes optically thin $(\tau\sim 1)$ and they explain the higher value of $\Sigma_{\rm crit}^{-}$
for helium by its low opacity that must accordingly be compensated for by higher surface density to give $\kappa\Sigma \sim 1$.
However, as seen from Table \ref{tab:crit}, this explanation cannot be correct. As already noticed and explained by \citet[][see also \citet{Smak99}]{can_whee84}
the optical depth at $\Sigma_{\rm crit}^{-}$ is \textsl{always} (for the standard assumption $\alpha \leq 1$)
larger than $1$ and the turnover at $\Sigma_{\rm crit}^{-}$  is related to the optical depth in a less straightforward way.

From Table \ref{tab:crit} one can see that at critical points the relation Eq. (\ref{eq:radeq})
is approximately satisfied only for $Y=1$. For the other two cases the temperature gradient is much
flatter than predicted by Eq. (\ref{eq:radeq}), reflecting the importance of convective energy transport,
consistent with the high optical depths of these configurations. Therefore for $Y=1$ (low opacity) the
turnover at $\Sigma_{\rm crit}^{-}$ results from the change in opacity temperature-dependence, whereas
for the high optical depth cases of $Z=0.02$ and $Z=0.04$ the change of slope results from the
(opacity-related) change in importance of convective energy transport.

Both (maximum and minimum) critical surface-density values are higher in helium dominated discs
than in discs where hydrogen dominates. 
\begin{figure}
\includegraphics[scale=0.9]{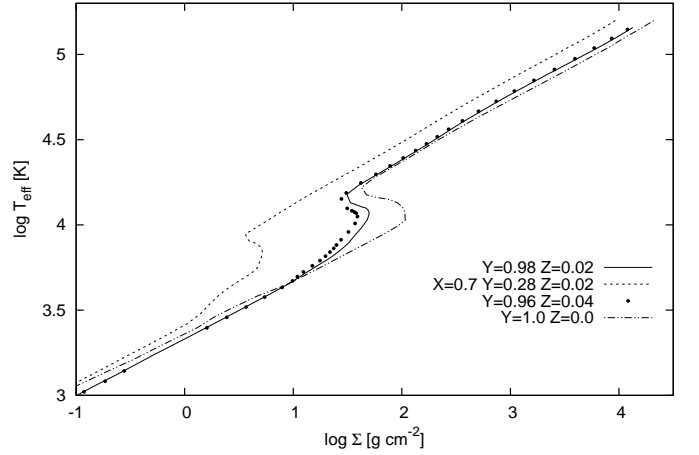}
{\caption{{\footnotesize S-curves for $M_1=1.0$, $R=10^{9}\,\,{\rm cm}$ , $\alpha=0.1$,
and four chemical compositions.}}
\label{fig:scurve}
}
\end{figure}
\begin{figure}
\begin{centering}
\includegraphics[scale=0.85]{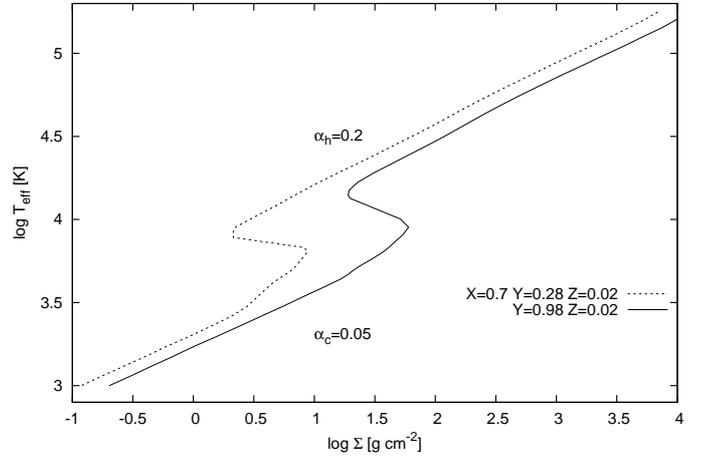}
\caption{{\footnotesize ``Effective" S-curves for combined cold and hot branches
corresponding to two values of $\alpha$: $\alpha_c=0.05$ and $\alpha_h=0.2$. 
The curve on the left corresponds to solar composition, while that on the right corresponds to $Y=0.98$.
$M_1=1.0$, $R=1.0\times 10^{9}\,\,{\rm cm}$.
}}
\label{fig:scurveff}
\end{centering}
\end{figure}
\renewcommand{\arraystretch}{1.2}\begin{table}
\begin{centering}
\begin{tabular}{|c|c|c|c|c|}
\hline 
 & ${\rm Y}=1.0$ & ${\rm Y}=0.98$ & ${\rm Y}=0.96$ & solar\tabularnewline
\hline 
\hline 
$\Sigma_{{\rm crit}}^{+}\,[{\rm g}/{\rm cm}^{2}]$ & $42.1$ & $30.8$ & $27.8$ & $3.5$\tabularnewline
\hline 
$\Sigma_{{\rm crit}}^{-}\,[{\rm g}/{\rm cm}^{2}]$ & $108.0$ & $50.2$ & $39.0$ & $5.2$\tabularnewline
\hline 
$T_{{\rm eff}}^{-}\,[{\rm K}]$ & $10700$ & $11700$ & $11200$ & $6960$\tabularnewline
\hline 
$T_{{\rm eff}}^{+}\,[{\rm K}]$ & $15600$ & $14250$ & $14200$ & $8210$\tabularnewline
\hline 
$\kappa(\Sigma,T_{c}^{-})\,[{\rm cm}^{2}/{\rm g}]$ & $4.2\times10^{-1}$ & $4.5\times10^{2}$ & $5.4\times10^{2}$ & $1.3\times10^{3}$\tabularnewline
\hline 
$\tau_{{\rm tot}}(\Sigma,T_{c}^{-})$ & $45.3$ & $2.24\times10^{4}$ & $2.11\times10^{4}$ & $7.03\times10^{3}$\tabularnewline
\hline 
\end{tabular}
\par\end{centering}
\centering{}
\caption{{\footnotesize S-curve critical points for four abundances. ${\rm R}=1.0\times10^{9}\,{\rm cm}$, $\alpha=0.1$}}
\label{tab:crit}
\end{table}
However, as has been realized by \citet{Smak84}, the S-curves used in the DIM must be modified
if the model is to reproduce the outburst properties of dwarf novae. The viscosity parameter $\alpha$ cannot
be constant during outburst, it must be larger in hot discs than in cold, quiescent discs. The DIM uses
``effective" S-curves by combining configurations with two different values of $\alpha$. Observations of  dwarf
nova normal outbursts imply that the ratio between the two is between 4 and 10. Examples
of effective S-curves are shown in Fig. \ref{fig:scurveff} for an $\alpha$ ratio equal 4.0.
This ratio has been assumed to be the same for a hydrogen- and helium-dominated disc. Since the $\alpha$
variation used is purely arbitrary, gauged through observed dwarf nova light curves, this assumption is
not necessarily valid. We will discuss this problem in more detail in Sect. \ref{sub:playalpha}.

\section{Mass transfer rates and stability}
\label{sect:mtranstab}

The first to consider the stability of helium accretion discs in AM~CVn systems was \citet{can84}. 
Here we use the latest Opal opacity tables and a larger statistics of AM~CVn stars to test the stability criteria
versus the observed properties of these systems. As discussed above, the critical parameters defining the disc's
stability depend on the chemical composition. 

\subsection{Chemical composition of accretion discs in AM~CVn stars.}

In general, observations clearly show that discs in AM~CVn
stars are not made of pure helium. We briefly describe the observed AM~CVn spectra to show what their disc metallicity might be.

The two persistently faint systems - GP~Com and V396 Hya - are very special. Unlike the other
(outbursting or persistently bright) AM~CVn stars, they show a
significant overabundance of NV, in addition to a strong HeI and a weaker HeII line, but not the Si line,
that is usually seen in other binaries of this type. This suggests metal poor
secondaries. Other binaries in a low state, such as SDSS~J0902 or SDSS~J1552 
also show weak SiII and FeII lines.

In the permanent high-state systems AM~CVn or HP~Lib the spectrum is
dominated by the absorption lines, but sometimes emission in HeII is
also detected. In the optical part of the spectrum the asymmetrical,
broad absorption line of HeI dominates. The UV spectrum shows broad
absorption lines of HeII, NV, NIV, SiIV, CIV \citep[][]{Wade07}.
The spectra of the erupting systems, CR Boo and V803 Cen have similar characteristics during outbursts.

For GP Com  \citet{Stro2004} estimated
$X_{\rm He}=0.99$ and $Z=0.01$. Because GP Com and V396 Hya are the two metal-poor systems, we can consider $Z=0.01$ as lower
limit for an AM~CVn disc metallicity. However, the exact metallicities of particular
systems have not been well determined yet \citep[][and private communication]{Neletal10}.

Throughout, our fiducial model of outbursting AM~CVn stars assumes $Y=0.98,\,Z=0.02$, but
when discussing the general properties of the systems and the model we will also consider other
possibilities. 

\subsection{Stability of observed systems}

In the standard version of the DIM the mass transfer rate $\dot{M}_{\rm tr}$ from the secondary
is assumed to be constant. Its value determines the stability of the disc.

For pure helium discs the critical values of the accretion rate are
\citep{LDK}
\begin{equation}
{\Mcrp}=1.01\times10^{17}\alpha_{0.1}^{-0.05}R_{10}^{2.68}M_{1}^{-0.89}\,\,{\rm g\,s^{-1}}
\label{eq:Mcrit_+}
\end{equation}
and
\begin{equation}
{\Mcrm}=3.17\times10^{16}\alpha_{0.1}^{-0.02}R_{10}^{2.66}M_{1}^{-0.89}\,\,{\rm g\,s^{-1}}.
\label{eq:Mcrit_-}
\end{equation}
To be in a hot (cold) stable equilibrium the accretion rate in a disc must be  higher (lower) \textsl{everywhere}
than the corresponding critical $\Mdot$.
Therefore the stability conditions are
\begin{itemize}
\item a stationary ($\Mdot(R)=\rm const.$) disc is hot and stable when $\Mtr > {\Mcrp}\left(R_D\right)$;
\item a stationary ($\Mdot(R)=\rm const.$) disc is cold and stable when $\Mtr < {\Mcrm}\left(R_{\rm in}\right)$.
\label{stab_cond}
\end{itemize}
The critical rate ${\Mcrm}$ for helium is 12 times higher than for hydrogen-dominated discs consequently the existence of
cold stable hydrogen-deficient discs does not require ridiculously low mass-transfer rates as in the case of  hydrogen-rich CVs, where it has to be lower than $8\times 10^{12}\,{\rm g/s}$.
The outer disc radius, $R_{D}$, is determined from the equation
\be
\frac{R_D}{a}=\frac{0.6}{(1+q)^{1/3}},
\label{Rd}
\ee
\citep[][]{Warner03}, where $a$ is the binary separation and $q=M_2/M_1$ the mass ratio. Since in AM CVn stars  $0.0125 \leq q \leq 0.18$ the dependence of the maximal disc radius
on mass ratio is rather weak and can be safely neglected.

The inner radius, $R_{\rm in}$, is taken to be equal to the radius of
the central white dwarf, $R_{\rm in}=R_{WD}$, which is determined by
the white dwarf mass $M{}_{\rm WD}$ through the $M-R$ relation \citep[][]{nau72}. 
The dependence on $\alpha$ in equations (\ref{eq:Mcrit_+}) and (\ref{eq:Mcrit_-})
is clearly negligible.
\begin{figure}
\begin{centering}
\includegraphics[scale=0.85]{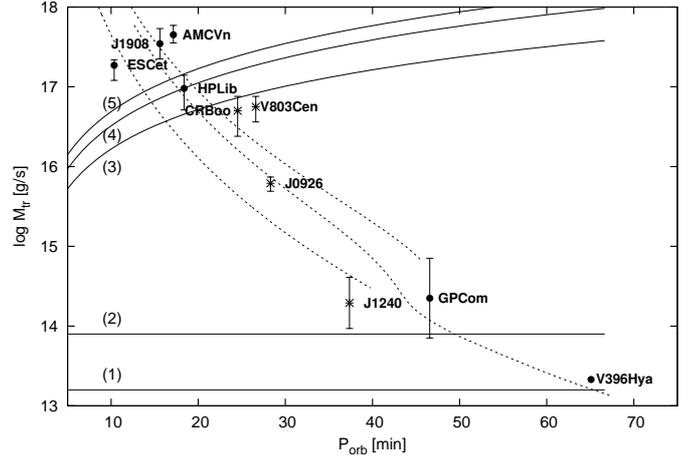}
\par\end{centering}
\caption{{\footnotesize The $\log \dot{M}_{\mathrm{tr}}$ -- $P_{\rm orb}$ plane. \textit{Dots} - persistent AM~CVn systems with known distances; \textit{asterisks} - outbursting AM~CVn systems with known distances; \textit{solid lines} - the upper and lower critical limits  $\dot{M}_{\mathrm{tr}}$ : (1) $\dot{M}^{-}_{\mathrm{crit}}$ for $M_{1}=1.0$, $\mathrm{Y}=1.0$; (2) $\dot{M}^{-}_{\mathrm{crit}}$ for $M_{1}=0.6$, $\mathrm{Y}=1.0$; (3) $\dot{M}^{+}_{\mathrm{crit}}$ for $M_{1}=1.0$, $\mathrm{Y}=0.96, \mathrm{Z}=0.04$; (4) $\dot{M}^{+}_{\mathrm{crit}}$ for $M_{1}=1.0$, $\mathrm{Y}=1.0$; (5) $\dot{M}^{+}_{\mathrm{crit}}$ for $M_{1}=0.6$, $\mathrm{Y}=1.0$.  ($M_1$ -- in solar units). \textit{Dotted lines}: Evolution models for AM~CVns through the WD channel (kindly provided by Chris Deloye).}}
\label{fig:mdotp}
\end{figure}
The plot of the estimated $\dot{M}_{\mathrm{tr}}$ versus $P_{\rm orb}$ for observed
AM~CVns (see Table \ref{tab:sys}) compared with the model predictions is shown in Fig.\ref{fig:mdotp}.
In agreement with the model, the permanently high-state systems ES~Cet, SDSS~J1908+3940 and  AM~CVn lie well above the critical mass transfer limit. 
Another bright and steady system is  HP~Lib. Although it is very close to the critical line of $\dot{M}_{\rm crit}^{+}$, there still exists a certain set of $\dot{M}_{\rm crit}^{+}$ -- lines above which HP~Lib lies with reasonable parameters.
The mass-transfer rates of CR~Boo and V803~Cen are very close to the upper critical line, which might explain why these two outbursting systems have been  alternately and confusedly classified as analogues of ER~UMa, Z~Cam, and (not consistent with their mass-transfer rate) of WZ~Sge stars. Below we test the hypothesis that they are `superoutbursting" Z~Cam-type stars.
GP~Com and V396~Hya have shown no changes in their brightness till now, they are
considered to be low-state systems. From Eq. (\ref{eq:Mcrit_-}) one concludes that their primary masses
$M_{1}$ should lie in the range $0.6-1.0\,\Msun$ (the mass-transfer rate for GP Com is an upper limit, see Table \ref{tab:sys}). 
\begin{table*}[lt]
\begin{centering}
\renewcommand{\arraystretch}{1.5}
\begin{tabular}{|c|c|c|c|c|c|}
\hline
System & ${\rm P}_{{\rm orb}}\,({\rm min})$ & $M_1(\Msun)$ &  $M_2(\Msun)$ & $\log\dot{M}_{tr}$ {[}g/s{]} & Ref.\tabularnewline
\hline
\hline
ES Cet$^\ast$ & $10.35$ & $0.44-0.69$ & $0.062 - 0.26$ & $17.27_{-0.19}^{+0.07}$ & $1,5$\tabularnewline
\hline
SDSS J1908+3940$^\ast$ & $\sim 15.6^\flat$ &    ?     &      ?          & $17.35 - 17.73$       &  7     \tabularnewline
\hline
AM~CVn$^\ast$ & $17.15$ & $0.71\pm 0.07$& $0.13\pm 0.01$ & $17.65_{-0.1}^{+0.12}$ & $2$\tabularnewline
\hline
HP Lib & $18.38$ & $0.80 - 0.4$ & $0.048 - 0.088$ & $16.98_{-0.27}^{+0.17}$ & $2$\tabularnewline
\hline
CR Boo & $24.52$ & $1.10 - 0.67$ & $0.048 - 0.088$ & $16.7_{-0.32}^{+0.18}$ & $2$\tabularnewline
\hline
V803 Cen & $26.6$ &$1.17-0.78$ & $0.059-0.109$ & $16.75_{-0.19}^{+0.13}$ & $2$\tabularnewline
\hline
SDSS J0926+3624 & $28.3$& $0.85 \pm 0.04$ & $0.035 \pm 0.003$& $15.79_{-0.1}^{+0.08}$ & $4,6$\tabularnewline
\hline
SDSS J1240-01 & $37.36$ & ? & ? & $14.29_{-0.32}^{+0.32}$ & $3$\tabularnewline
\hline
GP Com$^\ast$ &  $46.57$ & $0.50-0.68$ &$0.009-0.012$ & $\leq 14.35_{-0.5}^{+0.5}{^\dag}$ & $2$\tabularnewline
\hline
V396 Hya$^\ast$ &  $65.1$ & ? & ? & $13.33$ & $3$
\tabularnewline
\hline
\end{tabular}
\par
\end{centering}
\smallskip
\caption{\footnotesize Properties of AM~CVns with known orbital periods and distances. $^\ast$- persistent systems; $^\dag$ - upper limit, $^\flat$ - superhump
period~(?). \,References:\,
(1) \citet{Esp05}; (2) \citet{Roel07a}; (3) \citet{Bild06}; (4) \citet{Deloy07};
(5) \citet{Copper11b}; (6) \citet{Copper11a}. (7) \citet{Fontaine11}. See also \citet{NelemAM}.
}
\label{tab:sys}
\end{table*}

\section{Modelling outbursts of helium-dominated accretion discs}
\label{sect:models}

\subsection{The front structure}

The structure of heating and cooling fronts is affected by the atomic structure of the dominant
elements. Helium has two ionization energy levels: ${E}_{\rm HeII}=24.6\,\,{\rm eV}$
which corresponds to temperature ${T}_{\rm HeII}\sim28500\,{\rm K}$
for the ionization of the first electron, and ${E}_{\rm HeIII}=54.4\,\,{\rm eV}$,
i.e. ${T}_{\rm HeIII}\sim63000\,{\rm K}$, for the ionization of the second
electron. The existence of two electrons in the helium atom also opens 
a wide range of states where one electron is exited and the second
one is ionized, or both are excited to various energy levels.

For solar composition discs, a detailed analysis of the cooling and heating fronts has been performed by \citet{MHS}. 
They found that the form of the fronts does not depend on viscosity  and
front location in the disc. In hydrogen-dominated discs the temperature gradients of both heating
and cooling front are quite smooth (the bottom row of Fig. \ref{fig:profiles} shows surface density (left) and central temperature (right) profiles in solar-composition disc), showing no clear breaks.
\begin{figure}
\includegraphics[scale=0.7]{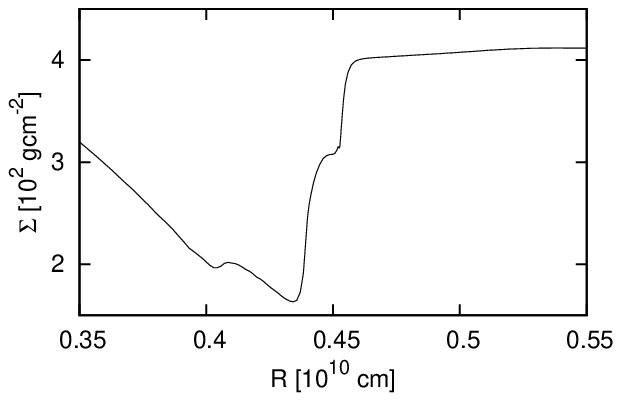}
\includegraphics[scale=0.7]{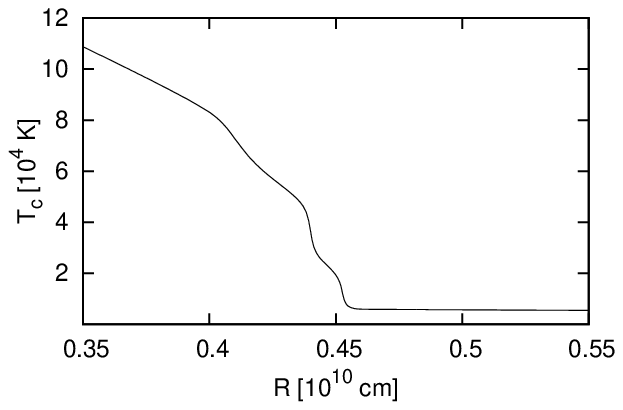}
\includegraphics[scale=0.7]{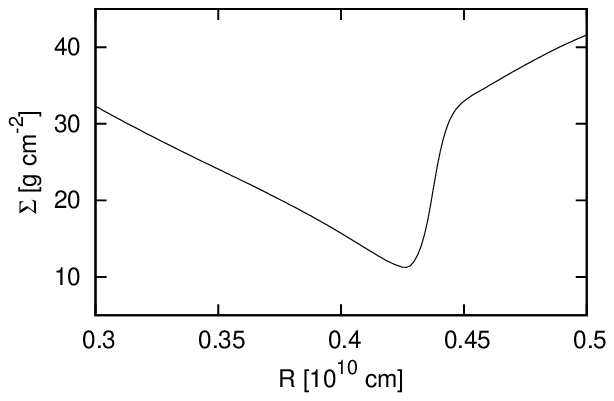}
\includegraphics[scale=0.7]{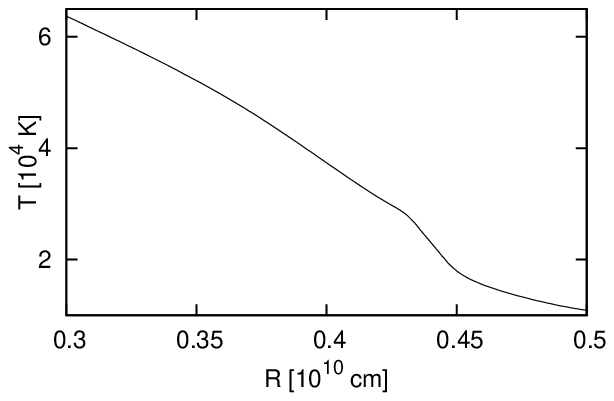}
\caption{{\footnotesize Cooling front structure: {\textit {top, left}}: Y=1, $\Sigma$ profile; {\textit {top, right}}: Y=1, $T_{c}$ profile; {\textit {bottom, left}}: solar, $\Sigma$ profile; {\textit {bottom, right}}: solar, $T_{c}$ profile. Parameters used for calculation are $\alpha_c=0.05$, $\alpha_h=0.2$, $M_1=1.0$, $\Mtr=1.0\times 10^{15}$ g/s, $\langle R_D \rangle=1.2\times 10^{10}$ cm.}}
\label{fig:profiles}
\end{figure}
In comparison, fronts in helium discs exhibit additional features (top row of Fig.\ref {fig:profiles}).
The breaks in the slope of the temperature gradient appear
at roughly the same temperatures for the heating and cooling fronts.
One can identify these temperatures with atomic transitions
in helium. The first break appears for ${T}=80000-92000\,\,{\rm K}$
($69-80\,\,{\rm eV}$) and can be assigned to double-photoionization and
different negative-ion resonances. The break at ${T}\sim50000\,\,{\rm K}$
($\sim42\,\,{\rm eV}$) corresponds to the state in which the atom is singly
ionized and the second electron is in excited state ${n}=2$.
In the region of temperatures around  ${T}\sim28000\,\,{\rm K}$
($\sim24\,\,{\rm eV}$) the neutral helium becomes singly ionized.
Below this is a range of temperatures at which one electron remains
in the ground state while second one is in the excited state. Other
points marked on $T$ and $\Sigma$ profiles are hard to classify
unambiguously. 

Yet more changes with the addition of  metals to helium. We consider
here discs with ${Y}=0.98$, ${Z}=0.02$ and ${Y}=0.96$, ${Z}=0.04$.
At the temperatures available in these discs, elements such as oxygen
or carbon become highly ionized. Despite low number fractions,
their contribution to the population of free electrons, and in turn their
impact on the opacities in the disc, is significant. For instance,
the fourth ionization energy of ${\rm O}$ is ${E}_{{\rm OV}}=77.7\,\,{\rm eV}$
(${T}\sim82000\,\,{\rm K}$), the fourth ionization
energy of ${\rm C}$ is ${E}_{{\rm CV}}=64.7\,\,{\rm eV}$ (${T}\sim68000\,\,{\rm K}$) and that of N (thought to be enriched in some AM CVn stars) is
$E_{\rm NV}=77.47 \,\,{\rm eV}$ $(T\sim \,\, 81 760 \rm K$).
Because the main opacity sources in the disc  are free-free and bound-free transitions (electron scattering, however,
is negligible), the additional electrons change the efficiency of the cooling and heating mechanisms. This
influences the maximum and minimum temperatures in the disc (see Table \ref{tab:crit}).
N enrichment does not contribute more than other metals
to the opacity growth.
For solar-composition discs we do not see the features from
atomic transitions of helium and metals, because fronts there start
to propagate at lower temperature and much lower densities (i.e. hydrogen
ionization temperature $\sim3.5\times10^{4}\,\,{\rm K}$ and $\Sigma\sim30\,\,{\rm g}\,{\rm cm}^{-2}$)
at which these transition do not take place or are negligible because
of the small contribution of metals to the overall disc composition (such as
single ionization of carbon at $T\sim12000\,\,{\rm K}$ ). 

The influence of the chemical composition on the details of the front structure does not manifest itself in the observational outburst cycle but helps understanding
the DIM physics.

\subsection{Playing with {\large $\alpha's$}}
\label{sub:playalpha}

The critical surface densities depend on $\alpha$ to the power $\sim 0.85$ so that an $\alpha$'s ratio of 4, say,
corresponds to $\Sigma$'s ratio of $\sim 3.2$. From Table \ref{tab:crit} we can see that this is just the
ratio of critical densities for a pure helium disc with $\alpha_{h}=\alpha_{c}$. Indeed, for such a disc the DIM produces outbursts
with amplitudes up to $\sim2.5\,\,{\rm mag}$ with $\alpha$ kept constant (Fig. \ref{fig:alpha1}).
However, this cannot correspond to real AM~CVn outbursts since, as mentioned above, observations show
that their discs also contain  certain amount of metals. We tested
our model for discs with metallicity $Z=0.0003,\,\,0.02$, and $0.04$.
For $Z\geq2\%$ one cannot obtain amplitudes larger than $1\,\,{\rm mag}$
without changing $\alpha$. Adding just 2\% of metals lowers
the amplitude to less than $1\,\mathrm{mag}$  (Fig. \ref{fig:alpha1}). Nevertheless, as we will below, in AM CVn  the required $\alpha$ jump can be lower than in dwarf novae (see e.g. Fig. \ref{fig:norm}).
\begin{figure}
\begin{centering}
\includegraphics[scale=0.9]
{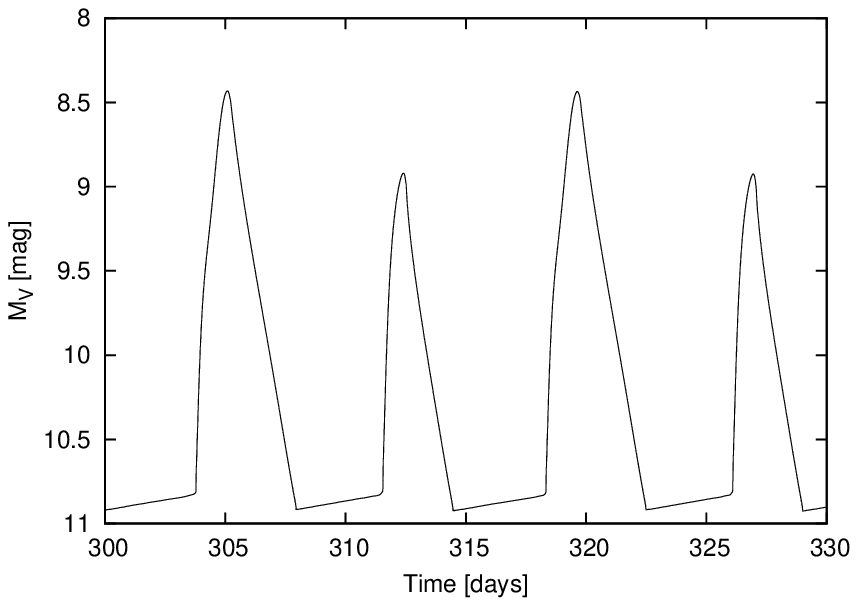}
\includegraphics[scale=0.9]
{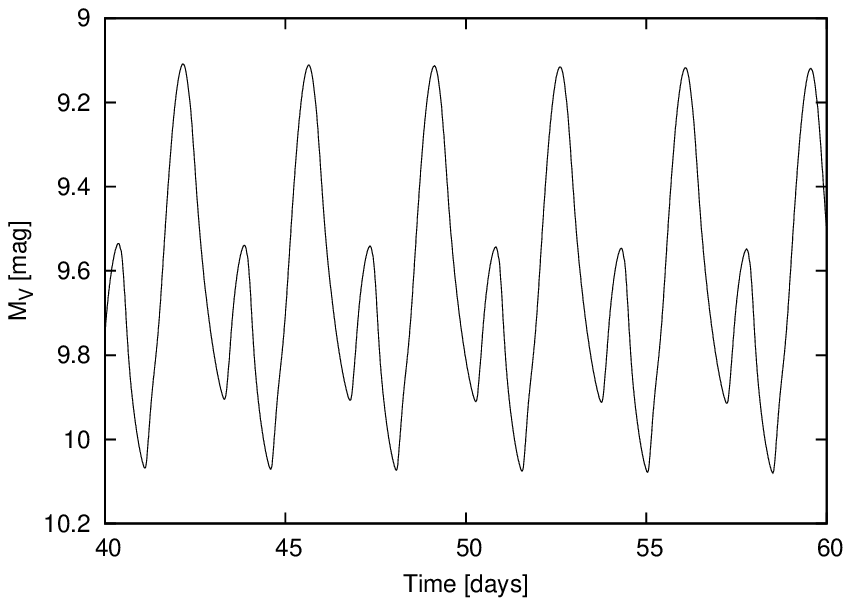}
\end{centering}
\caption{{\footnotesize Light curves calculated within the standard DIM with the same $\alpha$ in the hot and cold state for the discs with two different chemical compositions. \textit{Top}: Y=1.0. \textit{Bottom}: Y=0.98, Z=0.02. Parameters in
both cases : $\alpha_{h}=\alpha_{c}=0.1$, $\dot{{\rm M}}_{\rm tr}=10^{16}\,\,{\rm g/s}$, $M_1=1.0$ and the mean outer radius is $\langle {\rm R}_{D}\rangle=1.2\times10^{10}\,\,{\rm cm}$.}}
\label{fig:alpha1}
\end{figure}
One can notice also that changing the chemical composition modifies not only the amplitude but also
the shape of the outburst light-curve. This is to be expected because a diminished $\Sigma_{\mathrm{crit}}^-/\Sigma_{\mathrm{crit}}^+$
ratio favours the appearance of so-called ``reflares" \citep[see e.g.][]{Menou00,dubetal-01,L01}.

The accretion rate at outburst maximum is $\dot M_{\mathrm{peak}}\sim \left(\ah/\ac\right)^{8/7}$ \citep[see e.g.][]{dubetal-01}, hence the necessity of increasing the ratio of the viscosity parameters. There is not much to guide us in performing the change of $\alpha$. Numerical simulations of the magneto-rotational instability (MRI) which is believed to drive turbulence in Keplerian accretion discs \citep[see][and references therein]{Balbus03,Balbus05}, provide no guide in this case. Indeed, the value of $\alpha$ they produce is an order of magnitude lower than  required by observations \citep[e.g.][]{KPL07}. Transients appearing in some MRI simulations produce a viscosity coefficient of the correct order of
magnitude but they are too short-lived  to be the solution of this fundamental problem \citep[][and Reynolds, private communication]{Sortahiaetal11}.

The choice of the formula matching the two $\alpha$s is not without consequences because it can affect the properties of the outburst light-curves \citep[see e.g.][]{Cannetal11}. In our code we use the following interpolation formula:
\begin{equation}
\log\alpha(T_c) = \log\ac  + \left[\log\ah - \log\ac \right]\times \left[1 + \left(\frac{T_0}{T_c}\right)^8\right]^{-1}.
\label{eq:fitalpha}
\end{equation}
The motivation behind this expression is the wish to keep the values of $\Sigma^-(\alpha)$ and $\Sigma^+(\alpha)$ in the effective S curve
equal to the ``original" $\Sigma_{\mathrm{crit}}^-$ and $\Sigma_{\mathrm{crit}}^+ $ (H98). Much depends, however, on the choice of $T_0$.
H98 chose $T_0\approx T_c(\Sigma_{\mathrm{crit}})$ while \citet{dubetal-01} used $T_0=0.5 [T_c(\Sigma_{\mathrm{crit}}^-)
+ T_c(\Sigma_{\mathrm{crit}}^+)]$, which is more convenient for the irradiated disc they were considering (irradiation changes the critical values on the hot branch).
In the present paper we are using the H98 prescription.

\subsection{Additional sources of disc heating}

Additional sources of disc heating are known to be important in H-rich CVs, and we need to test their influence on the modelled light-curves for helium discs.

\subsubsection{Outer disc heating}

As the mass is transferred from the secondary, its stream hits the outer ring of the accretion disc, forming the so-called hot spot. In the impact region the disc annulus of width $\Delta R_{hs}$  is heated at a rate $Q_{i}$  with efficiency $\eta_{i}$  \citep{Buat01}: 
\begin{equation}
Q_{i}(R)=\eta_{i}\frac{GM_{1}\Mtr}{2R_{D}}\frac{1}{2\pi R_{D}\Delta R_{hs}}{\rm exp}\biggl(-\frac{R_{D}-R}{\Delta R_{hs}}\biggr),
\label{eq:Qi}
\end{equation}
where $M_{1}$  is the mass of the primary white dwarf, $\Mtr$\ is the mass transfer rate from the secondary and $R_{D}$  is the outer radius of the disc.

When the system is in the high state most of its luminosity comes from the hot accretion disc, especially from the inner parts through dissipation in the boundary layer, but during the low state as much as half of the visible light can come from the hot spot \citep{Smak10}. The magnitude of this contribution depends on the inclination of the disc: lower inclination means higher luminosity from the impact region. 

When the term $Q_{i}(R)$  is taken into account in energy conservation equation, the heating from the hot spot reduces the critical values of $\Sigma$, which facilitates triggering the outside-in outbursts for mass transfer rates lower than in absence of heatings. It also decreases the lower limit of the $\dot{M}_{{\rm tr}}$  range for which the disc should be unstable, and lifts the quiescence level. In several cases including this term in the equations improves the description of dwarf nova outbursts  considerably \citep[see e.g.][]{Buat01,Buat01b}. On the other hand, \citet{Smak02} expressed doubts concerning the relevance of hot-spot heating, noticing that most of the impact energy is radiated away at the hot spot location.

Another presumed source of the outer disc heating is the action of tidal stresses \citep[see e.g.][]{Buat01}. In this case \citet{Smak02} argues that effect is important only during outburst and limited strictly to the outer disc edge.

The two outer disc heating contributions are of the same order of magnitude. In the following, when testing the DIM's properties, we will use only hot-spot heating (Eq.\ref{eq:Qi}) because it represents the presumed effect on the outer disc structure.

\subsubsection{Irradiation by the hot white dwarf}

As shown by \citet{H99}, in dwarf novae irradiation by the hot white dwarf can seriously affect outburst structure and evolution.  In addition to stabilizing the innermost disc regions by maintaining their temperature above the ionization temperature, this irradiation has a {\sl destabilizing} effect on the adjacent parts of the flow. There, irradiation reduces the vertical disc temperature gradient which decreases the critical surface density $\Sigma_{\mathrm{crit}}^-$. In hydrogen-dominated discs $T_c^- \sim 8000$K while the  accreting white dwarf temperature reaches values from 15 000 to 50 000 K \citep{Sion07}. Effective temperatures of accreting white dwarfs in AM~CVns are expected to be in the same range \citep{Bild06}, and apparently they are \citep{Sion11}. However, the critical midplane temperatures of helium-dominated discs are in the range $T_c^- \sim 18000\,-\, 25 000 $K which, in general, reduces white dwarf irradiation to an unimportant effect that will be neglected below.

\subsection{Magnetic field of the primary}

Observations of SDSS~J080449.49+161624.8, whose optical spectra have properties similar to those of the hydrogen-rich intermediate polars, suggest that the primary's magnetic field also plays a role in AM~CVn stars \citep{Roel09}. If this is the case, the white dwarf magnetic moment will have a non-negligible influence on the inner part of the disc.  The magnetic pressure increases steeply with decreasing radius ($P_{\rm mag}\sim R^{-6}$ ) and in quiescence it might exceed the gas and ram pressures of the infalling matter up to radius $R_{M}$, disrupting the matter flow. During the outburst the mass accretion rate sharply rises and the ram pressure of matter dominates over the magnetic pressure so thatthe inner edge of the disc approaches the surface of the white dwarf. This mechanism is summarized in the formula for the inner disc radius \citep{fkr}:
\begin{equation}
R_{\rm in}=R_{M}=9.8\times10^{8}\dot{M}_{15}^{-2/7}M_{1}^{-1/7}\mu_{30}^{4/7} {\rm cm}, 
\label{eq:Rmag}
\end{equation}
where $\mu_{30}$  is the magnetic moment in units of $10^{30}\,{\rm G}\,{\rm cm}^{3}$, $M_{1}$ is the mass of the primary in solar masses, and $\dot{M}_{15}$  is the mass accretion rate in units of $10^{15}\,{\rm g}\,{\rm s}^{-1}$ .

Including formula (\ref{eq:Rmag}) in the model prolongs the quiescence time and outburst duration. The explanation is the same for small $M_{1}$: when $R_{{\rm in}}$  is larger, more mass has to be accumulated to cross $\Sigma_{{\rm crit}}^{-}(R)$, therefore, the quiescence time lengthens. But because more mass is now in the disc, its accretion and the decay from the outburst will take longer. The magnetic field does not influence the outburst amplitude.

\section{Normal outbursts of AM~CVn stars}
\label{sect:modellc}

From the theoretical point of view, normal outbursts of accretion discs are best defined as outbursts in which only the thermal-viscous instability is at play with no significant mass-transfer variations and no decisive role of the tidal interactions \citep[see][for a detailed description of normal outbursts]{L01}. 

As mentioned before, there exists a clear difficulty with testing the (helium) DIM on AM~CVn outbursts since, in these systems, most of the time it is not clear when and if an observed outburst is normal. \citet{Patt1997,Patt2000} proposed using the empirical relation between the observed outburst amplitude and the recurrence time (the ``Kukarkin-Parenago relation"; hereafter ``KP relation") as a test of ``normalcy" of outbursts.  To evaluate the relevance of this suggestion, we will now briefly discuss this empirical amplitude vs recurrence-time relation and its connection to the DIM \citep[see][for more details.]{KL1}

\subsection{The Kukarkin-Parenago relation for AM~CVn stars}
\label{KP}

\citet{KP34} suggested the existence of a relation between the amplitudes of cataclysmic-variable star outbursts and their recurrence times. As clarified by \citet{VP85} this correlation may indeed to be considered to exist if the sample considered is reduced to normal dwarf nova outbursts. Because of the large scatter of the observational data the form of the correlation is sample-dependent and uncertainties are rather large. For a selected subset of U~Gem type dwarf novae with well-measured amplitudes and recurrence times (see Table \ref{tab:UGem_KP})
the correlation has the form\footnote{For the sample in  \citet[][]{Warner03} the relation is
$A_{\rm n}=(0.7\pm0.43)+(1.9\pm0.22)\log{T_{\rm n}}$.}
\be
A_n=(1.3\pm 0.6)+(1.6\pm 0.3)\log T_n ,
\label{eq:KP1}
\ee
where $T_{{\rm n}}$ is measured in days and $A_{{\rm n}}$ in magnitudes. 
\begin{figure}[h]
\begin{centering}
\includegraphics[scale=0.9]{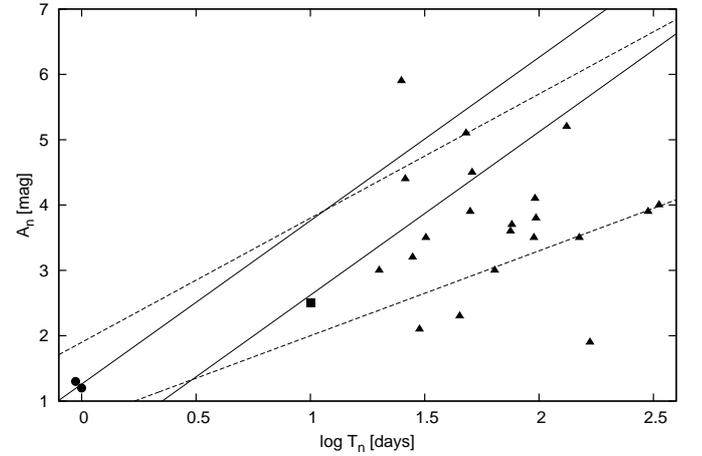}
\caption{{\footnotesize Amplitudes and recurrence times of normal dwarf nova outbursts and the Kukarkin-Parenago relation. \textit{triangles}: U~Gem stars; \textit{square}: PTF1~J0719+4858 (normal outburst); \textit{Circles}: V803~Cen and CR~Boo (cycling state rebrightenings). Data from \citet{Warner03} and \citet{Ritter03}. The dotted lines represent the upper and lower uncertainty of the observational KP relation fitted to the systems marked on the plot while the two solid lines represent the theoretical KP relation for helium discs - the lower line has been calculated for $\alpha_h=0.2$, $R_D=0.7\times 10^{10}$ cm, $M_1=1.0$ and the upper line for $\alpha_h=0.2$, $R_D=1.0\times 10^{10}$ cm, $M_1=1.2$. }}
\label{fig:KPobs}
\end{centering}
\end{figure}

The only AM~CVn system with unambiguous normal outbursts is PTF~1J0719+4858. 
Their amplitude is $A_{\rm n}\sim 2.5$ and recurrence time  $T_{n} \sim\,\ 10\, \rm{d}$.  Fig. \ref {fig:KPobs} shows the KP relation and the corresponding amplitudes and recurrence times of selected U~Gem stars (marked by triangles). The dotted lines represent the upper and lower uncertainty of the KP relation (Eq. \ref{eq:KP1}). Also plotted are the amplitudes and recurrence times for PTF~1J0719+4858 and the cycling states of the AM~CVn stars V803~Cen and CR~Boo. PTF~1J0719+4858 fits the KP relation well. The same is true, as remarked by \citet{Patt1997, Patt2000}, of the parameters of the cycling states of the two other AM~CVns: they are compatible with this relation. However, does thos imply that they can be classified as normal outbursts? This depends on what the KP relation is supposed to represent.

Using the standard version of the DIM, \citet{KL1} derived a theoretical $A_{\rm n}(T_{\rm n})$ relation in the form of
\begin{equation}
A_{\rm n}\approx C_{1}+2.5\log T_{\mathrm{n}},
\label{eq:An2}
\end{equation}
with
\begin{equation}
C_{1}\approx 1.5 - 2.5\log t_{\mathrm{dec}} +   BC_{\mathrm{max}}-BC_{\mathrm{min}},
\label{eq:C1}
\end{equation}
where $t_{\rm dec}$ is (as before) the outburst decay time calculated using the disc size and the viscous speed (see Eq. \ref{eq:vvist}) and $BC_{\mathrm{max}}$ and $BC_{\mathrm{min}}$ are the disc bolometric corrections in the high and low states, respectively. The viscosity parameter $\alpha_h$ must be in the range 0.1 -- 0.2 to correspond
to observations.

The relation Eqs. (\ref{eq:An2}), (\ref{eq:C1}) is based on the assumption that the whole mass accumulated during the time $t_{\mathrm{quiesc}}+t_{\mathrm{rise}}$ is accreted onto the white dwarf during outburst \citep[see][for details]{KL1}. Normal outbursts are supposed to satisfy this assumption therefore eruptions following the theoretical $A_n(T_n) $ relation can be considered to belong to this category.  In Fig. \ref{fig:KPobs} we plotted two theoretical $A_n(T_n) $ relations corresponding to helium discs with parameters $R_D=0.7\times 10^{10}$ cm, $M_1=1.0$ (PTF~1J0719+4858\footnote{The orbital parameters of this system are not yet known, however.}) and  $R_D=1.0\times 10^{10}$ cm, $M_1=1.2$ (V803~Cen and CR~Boo), respectively. In both cases $\alpha_h=0.2$. With these parameters the amplitudes and recurrence times of helium-dominated discs lie on the theoretical $A_n(T_n) $ relation.

\subsection{Model light-curves of normal outbursts}
\label{sect:lcmodels}

\begin{figure}[h]
\begin{centering}
\includegraphics[scale=0.9]{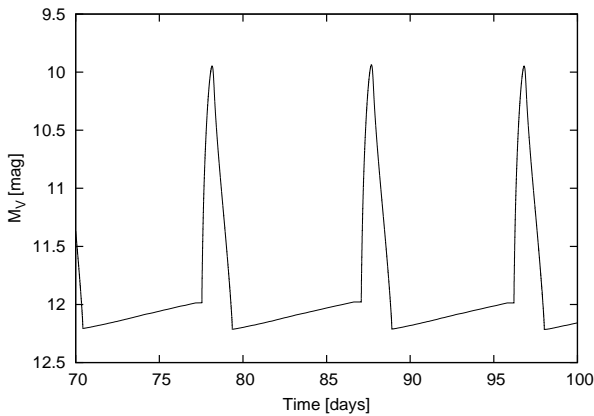}
\caption{{\footnotesize Normal outburst cycle for Y=0.98, Z=0.02 accretion disc. $\alpha_c=0.05, \alpha_h=0.1, \dot M_{16}=0.1, M_1=1.0, \langle R_D\rangle=1.1\times 10^{10}$ cm. The disc extends down to the white dwarf surface. No irradiation of disc or companion is taken into account.}}
\label{fig:norm}
\end{centering}
\end{figure}
In SU UMa stars the outbursts appearing between the superoutbursts are believed to be of normal type. The case of PTF~1J0719+4858, where a similar sequence is present, suggests that the same is probably true for AM~CVn stars. 

Fig. \ref{fig:norm} shows a synthetic light curve produced with the DIM applied to the helium-dominated discs. Although one does not expect to observe the light curve analogous to Fig.\ref{fig:norm} among compact helium binaries, the basic DIM should describe this part of the outburst cycle where normal outbursts are present. 

We briefly recall how the model's free parameters $\alpha_{h}$ , $\alpha_{c}$  and mass transfer rate $\dot{M}_{{\rm tr}}$ influence the recurrence times $T_{n}$ and amplitudes $A_{n}$  of normal outbursts. The primary mass $M_{1}$ , fixing the inner disc radius $R_{{\rm in}}$, is often difficult to estimate from observations and then it can be considered a free model parameter as well. The disc radius is constrained by observations, so there is not much freedom left in setting $R_{D}$  in the model, as long as one intends to refer to the real systems. 

The time elapsing between the onsets of two consecutive outbursts extends with increasing ratio $\beta$  of viscosity parameters: $\beta={\alpha_{h}}/{\alpha_{c}}$. Since the fiducial value of $\alpha_{h}$  is $0.1-0.2$, it is $\alpha_{c}$  which decides about $T_{n}$. However, $\beta$  also controls the value of $A_{n}$. Accordingly, the increase of $\beta$  (by lowering $\alpha_{c}$) not only results in longer recurrence times, but also in higher amplitudes. The outbursts will be more frequent in the disc fed with matter at high rates or/and in a system with heavier primary.

High $\dot{M}_{{\rm tr}}$  is also responsible for the appearance of wide normal outburst, which should not be confused with superoutbursts \citep[see][]{L01}.

Adding metals to helium increases $\Sigma_{\rm crit}^{-}$, due to the additional non-negligible contribution from highly ionized metals to the opacity. Furthermore, the consequence of augmented ${\rm Z}$  is the reduction of the total amount of helium and $\Sigma_{\rm crit}^{+}$  drop. The conclusion is that adding elements heavier than He to the disc matter is equivalent to diminishing the $\Sigma$ ratio (decreasing the $\beta$ ratio). This results in shorter recurrence times and lower outburst amplitudes.

It is worth emphasizing that metallicity affects the value of $\alpha_{c}$, which needs to be ``adjusted" to obtain the observed outbursts amplitudes and recurrence times. The difference between $\alpha_{h}$ and $\alpha_{c}$ to be explained by MRI may not be as large in a helium disc as usually thought for a solar-composition disc. However, a difference is still required, as argued in Sect. \ref{sub:playalpha}.

\section{AM~CVn outburst supercycles}
\label{sect:super}

Light curves of AM~CVn can be divided into three categories, which we refer to as: KL Dra - like, \ptf - like, and CR Boo/V803 Cen - like. The first category exhibits superoutbursts only and might be the equivalent of WZ Sge-type dwarf nova stars\footnote{KL~Dra might not be really of this type since small amplitude normal could have been missed, but an AM~CVn analog of the WZ~Sge-type presumably exists.}. \ptf \ almost certainly corresponds to the SU UMa type, well-known in hydrogen-dominated dwarf novae. The most mysterious are CR Boo and V803~Cen, which have been variedly described as equivalent to WZ~Sge, SU~UMa, ER~UMa, and Z~Cam stars.  It seems that they belong to all these classes partaking freely of such a wide range of dwarf nova behaviour, to quote \citet[][]{Patt2000}. 

As mentioned in Sect. \ref{sect:prelim}, in the present context we have chosen to opt for the EMT model as the model explaining the mechanism of superoutbursts appearance.

Although there is ample evidence for both the secondary's irradiation and the enhanced (during outburst) mass-transfer rate \citep[][]{Smak11}, the connection between the two is far from being clear. \citet{VH07,VH08} have shown that matter heated up by irradiation of the companion's disc-unshaded surface has cooled down too much by the time it arrives in the vicinity of the $L_1$ point to affect mass transfer towards the disc. Although the treatment of the cooling is simplified \citep{Smak09d}, it is unlikely that the correct treatment would modify the result significantly. A solution for this difficulty would be to have direct irradiation of the  $L_1$ region allowed by a tilted/warped disc  \citep[][]{Smak09e}.

Below, inspired by previous work, we will attempt to mimic the effect of irradiation on the mass-transfer rate by suitable parametrizations. 

\subsection{KL Dra and \ptf\,  -- like outbursts}

The general properties of the relatively simple light curves of KL Dra and \ptf \ can be reproduced by using the simple prescription for accretion-irradiation mass-transfer rate increase originally proposed by \citet{HLH}:
\begin{equation}
\dot{M}_{{\rm tr}}={\rm max}(\dot{M}_{0,{\rm tr}},\gamma\dot{M}_{{\rm acc}}),
\label{eq:gamma}
\end{equation}
where $0 \leq \gamma\leq 1$. $\dot{M}_{0,{\rm tr}}$ corresponds to the ``secular" (non-enhanced) mass-transfer rate while $\dot{M}_{{\rm acc}}$ is the accretion rate onto the primary.
\begin{figure}[h]
\includegraphics[scale=1.0]{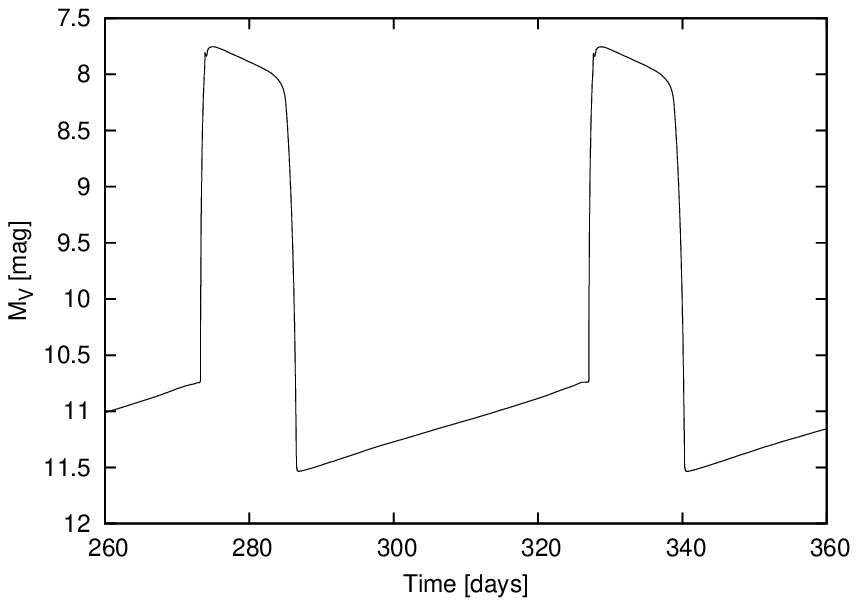}
\includegraphics[scale=0.68]{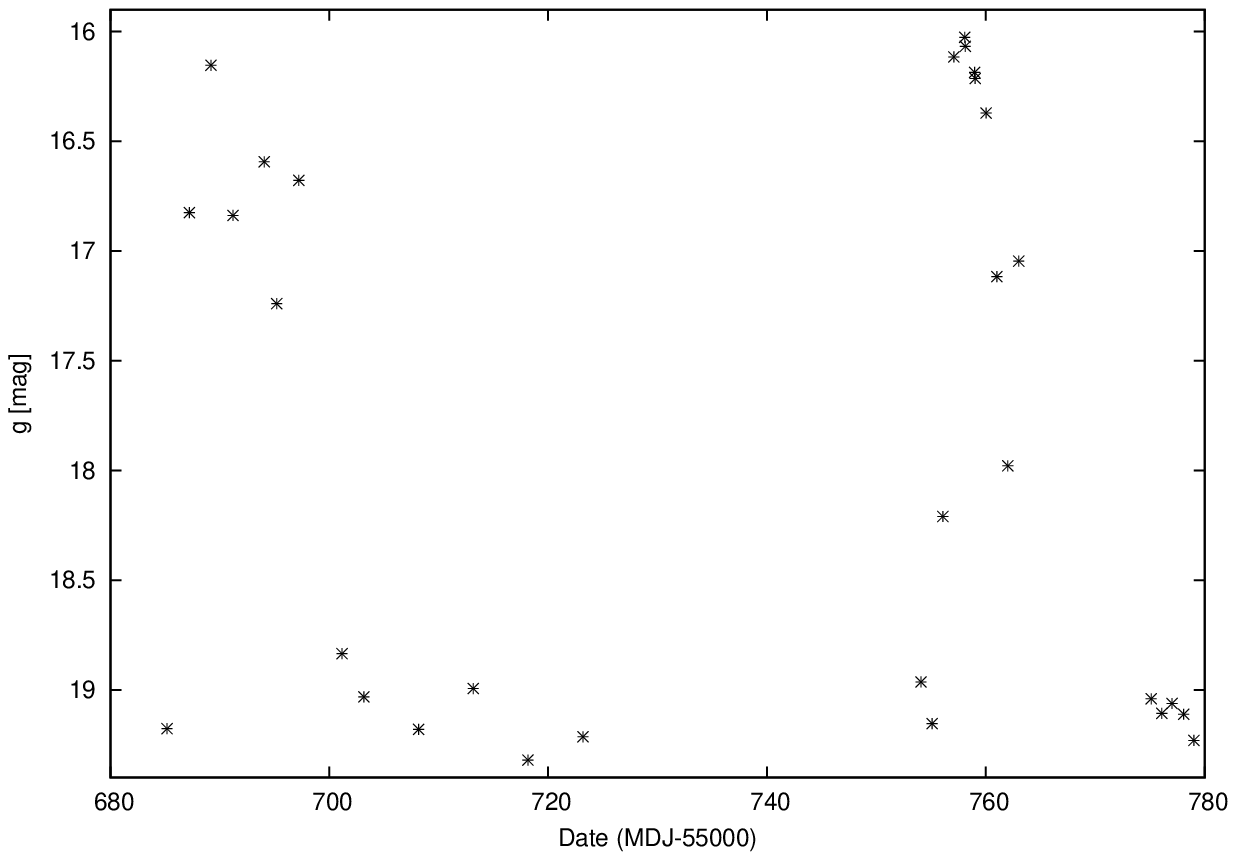}
\caption{{\footnotesize \textit {Top} : KL~Dra model: $\alpha_{c}=0.035$, $\alpha_{h}=0.2$, $\dot{M}_{{\rm tr}}=2\times10^{16}\,{\rm g/s}$, $M_{1}=0.6$, $\gamma=0.8$, $\langle {\rm R}_{D}\rangle=1.2\times 10^{10}\,\,{\rm cm}$, $\mu_{30}=1$. \textit {Bottom}: The part of KL~Dra light curve showing two subsequent superoutbursts. The time between their maxima is $~60$ days, their amplitude is $~3$ mag and their duration is $\sim 13$ days. Data provided by G. Ramsay.
}}
\label{fig:kldra}
\end{figure}

Dealing with better-sampled multi-wavelength light curves such as that of VY Hyi , \citet{SHL04} used a more ``refined" prescription using a suitably averaged over time $\dot{M}_{{\rm acc}}$ but using it here would be exaggerated. 

As for hydrogen-dominated dwarf novae \citep[][]{HLW00} the light curve properties depend on the mass-transfer rate, the mass of the primary, and the viscosity parameters, as well as on the assumed white dwarf's magnetic moment and the value of the parameter $\gamma$. In AM~CVns one has additionally to fix the chemical composition, i.e. the metallicity which we will assume to be $Z=0.02$ ($Y=0.98$) unless stated otherwise. We assumed $\langle {\rm R}_{D}\rangle=1.0\times 10^{10}\,\,{\rm cm}$ for the average size of the accretion disc.

\subsubsection{Superoutbursts only}

To obtain a light curve similar to that of KL Dra,  we used parameters close to those suggested by \citet{Rams10} for this binary:  $\dot{M}_{{\rm tr}}=2\times10^{16}\, \rm g\,s^{-1}$,  $M_{1}=0.6\Msun$. With the choice of the viscosity parameters $\alpha_{c}=0.035$, $\alpha_{h}=0.2$, choosing $\gamma=0.8$ and using  $\mu_{30}=1$ (corresponding to a  magnetic field $B\sim 1.5\times 10^3\, {\rm G}$) we obtained the supercycle shown in Fig. \ref{fig:kldra}. The calculated amplitude $A_{s}\approx 3$ mag, the (super)outburst duration $T_{\rm dur}\approx 15$ d and the recurrence time $T_{\rm recc}\approx 57$ d correspond quite well to the observed  $A_{s}\approx 3.5$ mag,  $T_{\rm dur}\approx 14$ d and $T_{\rm recc}\approx 63$ d.

Including the magnetic field of the primary and the heating by the hot spot was essential for obtaining the light curve in Fig. \ref{fig:kldra}. These two effects stabilize the disc.
The magnetic field truncates the inner parts of the disc, increasing $\Sigma_{\rm crit}^{-}(R_{\rm in})$. The disc then has to accumulate more mass to trigger the inside-out outburst.
The hot spot heating, in turn, lowers $\Sigma_{\rm crit}^{+}(R_{D})$ so that more mass has to be accreted before the cooling front can form. This favours triggering of  a superoutburst instead of a normal outburst. Both effects contribute to the suppression of normal outbursts. The resulting light curve compares well with the observations (Fig. \ref{fig:kldra}).

In reality, as mentioned above, KL~Dra could have normal outbursts that have been missed during observational campaigns. If true, this would mean that one should drop  from the disc evolution equations e.g. the hot-spot heating term to obtain normal outbursts.  This is what we have done in the next subsection when calculating models of the AM~CVn equivalents of SU~UMa stars.

\subsubsection{Helium ``SU UMa stars"}

Except for its orbital period ($\approx$27.7 min), not much is known about the parameters of the first observed helium SU UMa-type system \ptf. We assumed that this type of system transfers mass close to its secular mean taking for  example $\dot{M}_{{\rm tr}}=6\times10^{16}\,\rm g\,s^{-1}$. Assuming  $M_{1}=1.0\,\Msun$, $\alpha_{c}=0.02$, $\alpha_{h}=0.1$, $\gamma=0.6$ and a disc extending down to the white dwarf's surface, we obtain the light curve presented in Fig. \ref{fig:PTF}.

Between superoutbursts one notices a slight gradual increase of the minimum and maximum luminosity of the four consecutive normal outbursts.
This happens because during each normal outburst \Mtr \ is enhanced due to the $\gamma$-prescription - the disc gains more mass than it looses due to accretion. In consequence, during the sequence of the normal outbursts, the mass is gradually accumulated in the disc and $\Sigma$  rises everywhere in the disc. This effect is one of the typical features of the EMT model \citep[it is also present for slightly different reasons in the TTI version, see][]{SHL04,TO97}. While the general luminosity rise in quiescence is not observed in real systems and is one of the well-known weaknesses of the DIM \citep{Smak00}, the increase of the normal outburst peak luminosity might have been observed in some systems.

The superoutburst precursor is also typical of the EMT and TTI models \citep[][]{SHL04}. Owing to mass accumulation, the disc arrives to a state where after the rise of a normal outburst the cooling front is no longer able to propagate -- the disc becomes stuck in the hot state and a superoutburst begins. This last normal outburst leading to a superoutburst appears as its precursor in the light curve.

When one compares the model light curve to that of \ptf \ one notes that in our model the amplitude of normal outbursts is larger by $0.5\,{\rm mag}$ and their duration is four times longer. 
The normal outburst duration is extended by the \Mtr \ enhancement and the high amplitude is the result of the assumed $\alpha$ ratio -- a lower ratio would result in a lower amplitude. However, lowering this ratio would also lower the superoutburst amplitude and shorten the recurrence times. Tuning the parameters to obtain a better agreement does not make much sense in view of the arbitrariness of the mass-transfer prescription and of the uncertainties of the DIM itself \citep[for a discussion of the weaknesses of the DIM see e.g.][]{L01}.
\begin{figure}[t]
\includegraphics[scale=0.88]{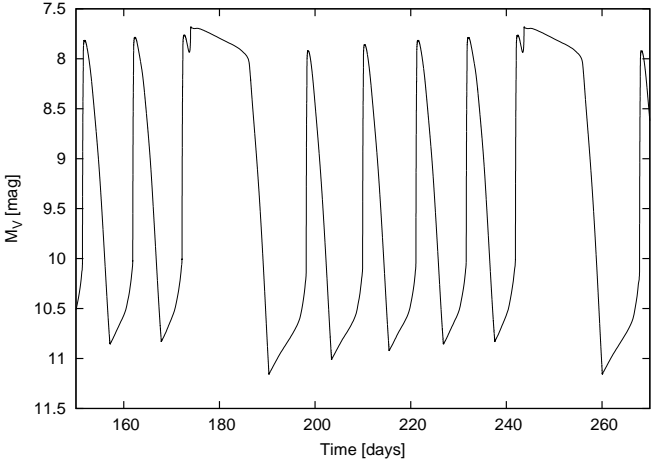}
\caption{{\footnotesize SU UMa type model for a helium-dominated disc (an ``imitation" of \ptf): $\alpha_{c}=0.02$, $\alpha_{h}=0.1$, $\dot{M}_{{\rm tr}}=6\times10^{16}$\,g/s, $M_{1}=1.0$, $\gamma=0.6$, $\langle {\rm R}_{D}\rangle=1.0\times 10^{10}\,\,{\rm cm}$, $\mu_{30}=0$.}}
\label{fig:PTF}
\end{figure}

\subsection{Dips, ``cycling", and standstills}
\begin{figure}[h]
\includegraphics[scale=0.8]{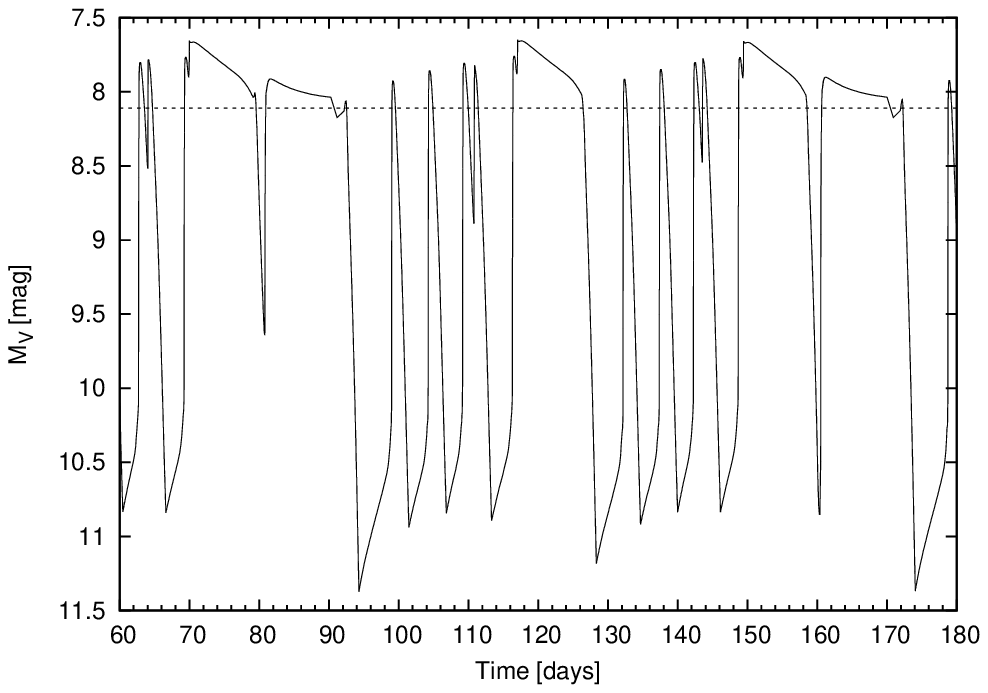}
\includegraphics[scale=0.8]{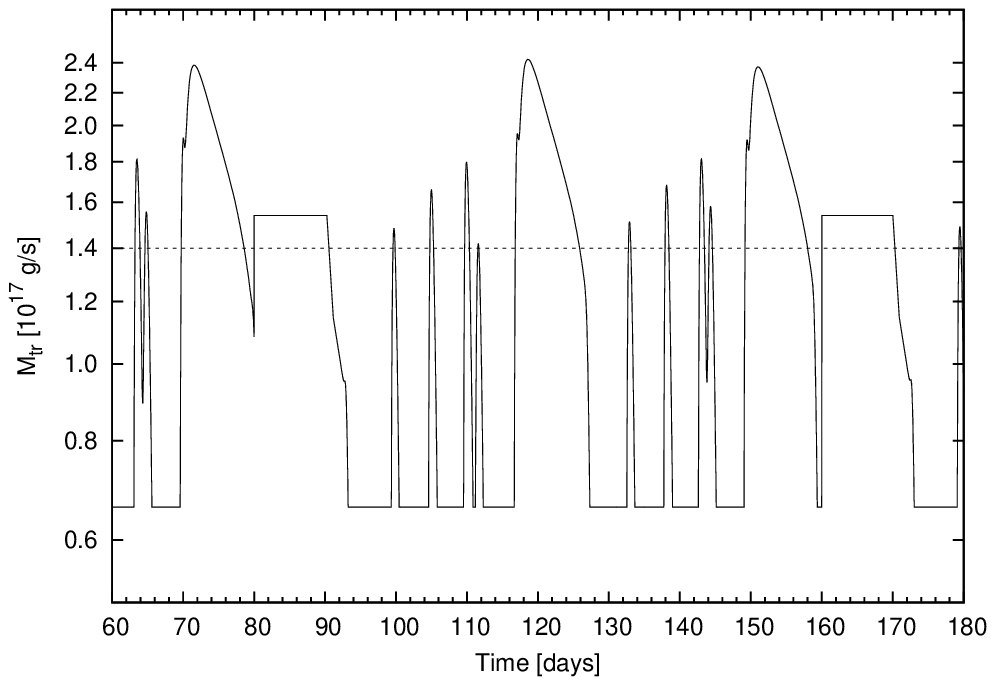}
{\caption{{{\footnotesize Light-curve of a superoutbursting helium ``Z~Cam" star. The parameters are: $\alpha_{c}=0.04$ , $\alpha_{h}=0.2$ , $M_{1}=1.0$, $ \dot{M}_{0,{\rm tr}} =1.1\times 10^{17}\,{\rm g/s}$, $\Delta{\dot M}_{\rm tr}=0.4\dot{M}_{0,{\rm tr}}$, $T_{\rm high}=10\,{\rm d}$, $T_{\rm low}=70\,{\rm d}$, $\gamma=0.7$, $\langle {\rm R}_{D}\rangle=1.0\times 10^{10}\,\,{\rm cm}$. The dotted lines correspond to the hot stability limit. {\textit {Top}}: the model lightcurve. {\textit {Bottom}}: \Mtr \ modulations. }}}
\label{fig:ZCam_He}
}
\end{figure}
Most (maybe all)  non-stationary AM~CVn stars exhibit at least one ``dip" during the decay from the superoutburst maximum. In some systems, such as  CR~Boo and V803~Cen,
the system becomes stuck in a state in which series of $\sim 1.0\,~{\rm mag}$ amplitude, $\sim 0.8-1.0\,{\rm d}$ recurrence-time outbursts are present. There is also some uncertainty about the nature of these intermissions in the superoutburst cycle: on the one hand, \citet{Patt1997} considered them to be a ``cycling state" of normal dwarf nova outbursts, on the other, \citet{Kato2001a} interpreted the same part of the (CR~Boo) light curve as a standstill (analogous to that of Z~Cam dwarf novae), arguing that no superoutbursts  are present during the high-luminosity state maintained by the system.

To answer which of these two interpretation is correct more observations are needed  and therefore we are not trying here to decide this question. Let us just note that the mass-transfer rates attributed to both CR~Boo and V803~Cen (see Table \ref{tab:sys}) are very close to the critical ones and a phenomenon analogous to Z~Cam standstills could indeed be expected for this systems. The main difference is that Z~Cam stars have no superoutbursts. We will therefore adapt the Z~Cam model \citep[][]{Buat01b} by combining its mass-transfer modulations with the EMT.

Dips might or might not be related to the cycling states. They are very similar to those observed during the decay from superoutbursts in WZ~Sge -- type stars. If common to both systems, the  dip origins are not connected to the value of the mass-transfer rate but would instead result from the very compact size of their orbit. We tested a simple hypothesis based on this assumption.

\subsubsection{``Z Cam -- type" modulations and ($\gamma$) superoutbursts}

Adapting the Z Cam models of \citet{Buat01b} to AM~CVn stars, we assumed the mass-transfer to be modulated as $\Delta \dot M/\langle \dot M \rangle=40\%$ around an average rate close to the critical ($\dot{M}_{{0,\rm tr}}=1.1\times10^{17}\,\rm g\,s^{-1}$) with the modulations occurring at $t=10 + k \cdot 70$ days. To take into account the presence of irradiation-induced super-outbursts, we combined this modulation with the mass-transfer modulations given by Eq. (\ref{eq:gamma}) with the actual value of the mass-transfer rate instead of $\dot{M}_{0,\rm tr}$. 

Choosing as typical parameters $\alpha_{c}=0.04$, $\alpha_{h}=0.2$, $M_{1}=1.0\,\Msun$, $\langle {\rm R}_{D}\rangle=1.0\times 10^{10}\,\,{\rm cm}$ and $\gamma=0.7$ we obtained the light-curve shown in Fig. \ref{fig:ZCam_He}. This light curve shows superoutbursts and standstills, as expected, but features resulting from the superposition of both types of modulations  are also present in the form of ``dips". The hypercycle (the cycle between two consecutive superoutbursts followed by a  standstill) starts with a major \Mtr \ enhancement due to $\gamma$ (about day $70$ on Fig.\ref{fig:ZCam_He}).  During the decay from the superoutburst maximum, the fall of \Mtr \ is quenched by a rising modulation through $\dot{M}_{0,{\rm tr}}+\Delta{\dot M}_{\rm tr}$. The mass-transfer rate then rises again (around day $80$) and the decay from maximum is reversed by a resulting outside-in heating front that catches up with the propagating cooling front before the latter reaches the inner disc edge -- hence a dip-like feature. During the following superoutburst a slight difference in phase of the two mass-transfer modulations allows the cooling front to propagate almost to the disc's inner end. A similar mechanism produces the dip in narrow outburst preceding the superoutburst. This suggest that dips and cycling-state features might result from mass-transfer modulations triggering cooling/heating front ``catching" and not from reflections. On the other hand, the indentation at the end of the standstill is the result of cooling front reflection \citep[as described e.g. in][]{dubetal-99}.

In any case, the presumed Z~Cam-effect mechanism applied to outbursting AM~CVns produced the required type of light-curve combining superoutbursts with standstills (and normal outbursts).

\subsubsection{Sinusoidal plus $\gamma$ modulation}

The prescription for mass-transfer modulations used in the previous section has been chosen not for its realism but because it has been used in modelling the Z Cam star outburst cycle by \citet{Buat01b}. Since in AM~CVn light curves other peculiar features are present in addition to standstills, we tried other simple forms of mass-transfer rate modulations.
The simplest is to modify Eq. (\ref{eq:gamma}) as
\begin{equation}
\dot{M}_{{\rm tr}}={\rm max(\dot{M}_{0,{\rm tr}}(1+A\,{\rm sin}(C+\pi t/\tau), (\gamma\dot{M}_{{\rm acc}})) ) },
\label{eq:MdotSin2}
\end{equation}
where $A$, $C$ (dimensionless) and $\tau$ (time) are adjustable constants; $t$ is the time coordinate. 

With  $M_{1}=1.0\,\Msun$, $\dot{M}_{{\rm tr}}=1.1\times10^{17} {\rm g/s}$ and $\langle {\rm R}_{D}\rangle= 1.0\times 10^{10}\,\,{\rm cm}$,  $\alpha_{c}=0.04$, $\alpha_{h}=0.2$, choosing $A=0.5$, $C=2$, $\tau=1\, {\rm d}$ and $\gamma=0.8$, one obtains the light-curve shown in Fig. \ref {fig:dipping}. One remarks superoutbursts with amplitude $A_{s}\sim 3\,{\rm mag}$, duration $T_{\rm s}\sim 15\,{\rm d}$ and recurrence time of $T_{\rm recc,s}\sim 27\,{\rm d}$. These are ``standard" EMT, $\gamma$ enhancement triggered outbursts.

In addition, short outbursts with a repetition time of $\sim 1.8\, {\rm d}$ and an amplitude $\sim 1.2-2.4\,{\rm mag}$ appear during the decay from superoutburst and persist until the following one. One is of course tempted to identify them with the cycling states of CR~Boo or V803~Cen. The origin of these short outbursts is quite simple to understand.
At the end of the decay of \Mtr \ from the superoutburst value, the sinusoidal enhancement of \Mtr \ dominates over the $\gamma$-induced \Mtr\ fluctuations. This short enhancement of \Mtr \ results in a low-amplitude, short outburst (starting at day $\sim 73$ in Fig.\ref{fig:dipping}) because the cooling front that switches off the superoutburst is quenched by a freshly triggered outside-in heating front. This happens because the post cooling-front $\Sigma$ is still close to the critical value due to \Mtr \ enhancement through the sinusoidal modulation.
However, since the disc's mass is decreasing, after one day a new cooling front forms immediately and this time it succeeds in propagating till the inner disc's rim but, because of mass accumulating, a new outburst starts immediately. All heating fronts are triggered by mass-transfer enhancement and therefore are of the outside-in type. Their maximum brightness is slightly increasing because of mass accumulation. Their amplitude depends on the combined strengths of the sinusoidal plus $\gamma$ enhancement. If the enhancement brings \Mtr \ above $\Mcrp$, the resulting short outburst will have a small amplitude as the outside-in heating front forms easily and promptly catches up with the cooling front that quenches it. But when the disc's mass is still quite low and the $\gamma+sine$ enhancement is not powerful enough to bring $\Mtr$ above $\Mcrp$, the cooling front will propagate unhindered till the low-luminosity level (see e.g. day $81$).
\begin{figure}[h]
\includegraphics[scale=0.8]{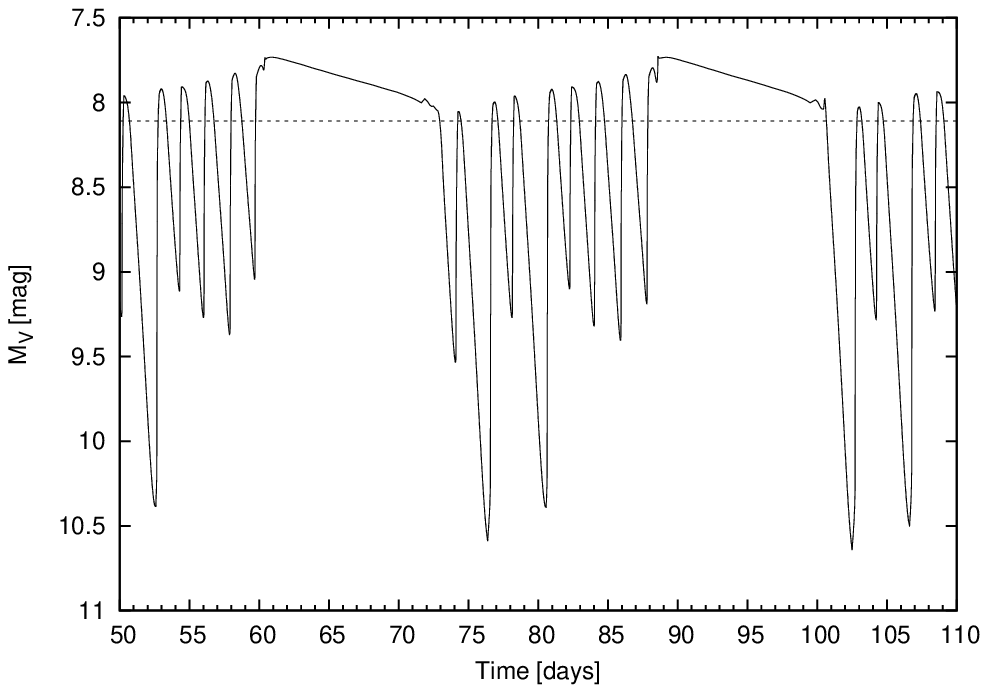}
\includegraphics[scale=0.8]{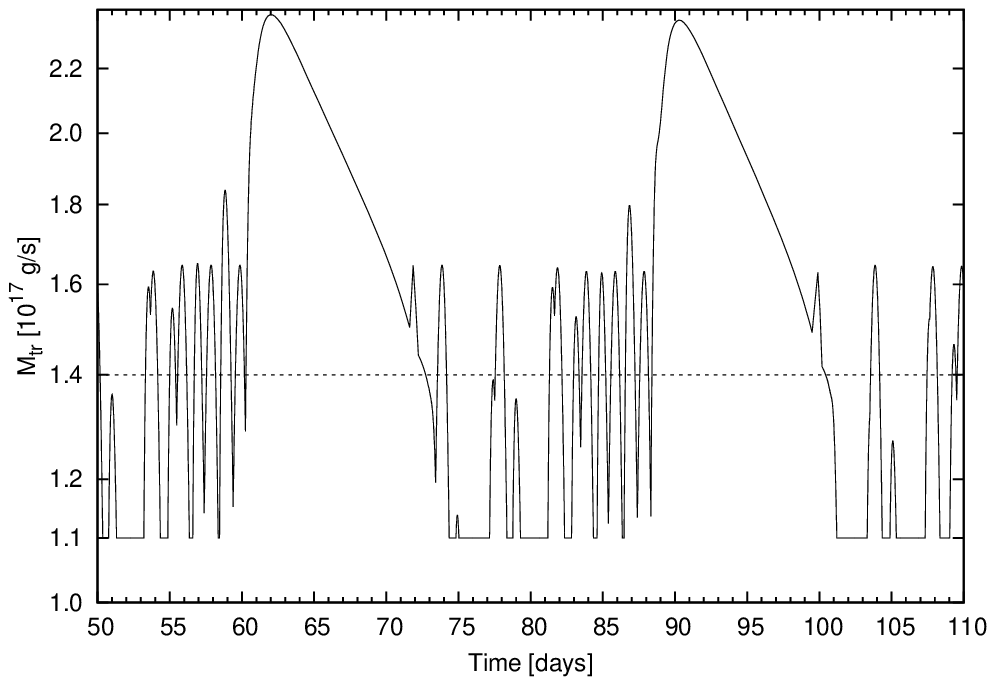}
\caption{{\footnotesize Light-curve corresponding to sinusoidal + $\gamma$ modulations of the mass-transfer rate. The parameters are $\alpha_{h}=0.2$, $\alpha_{c}=0.04$, $\dot{{\rm M}}_{0,\rm tr}=1.1\times 10^{17}\,\,{\rm g/s}$, $M_{1}=1.0$, $\langle {\rm R}_{D}\rangle=1.0\times 10^{10}\,\,{\rm cm}$, $\gamma=0.8$, $A=0.5$, $C=2$, $\tau=1$d. The dotted lines correspond to the hot stability limit. {\textit {Top}}: the model light-curve. {\textit {Bottom}}: \Mtr \ modulations. }} 
\label{fig:dipping}
\end{figure}
It seems therefore that suitable modulations of the mass-transfer rate can produce various observed features of outbursting AM~CV light curves even with the simplest assumptions about their shapes. However, the mechanisms that are able to produce mass-transfer modulations are still to be firmly identified. There is no doubt that mass-transfer rates in close binaries are highly variable, in some cases with huge amplitudes. Although irradiation of the secondary is the usual suspect, in many cases the observed variability \citep[e.g. of AM~Her or VY~Scl stars, see][]{Warner03} is not caused by irradiation. A periodic modulation of mass transfer rate could be also achieved by a warped/tilted disc \citep[][]{Smak11}. The sinusoidal variation  of Eq. (\ref{eq:MdotSin2}) could be considered an attempt to represent the effect of the variable irradiation if the modulation time $\tau =1$d could find a convincing interpretation in this context.

\section{Discussion and conclusion}
\label{sect:concl}

We have presented a systematic study of the dwarf nova disc instability model applied to outbursts of AM~CVn stars. What distinguishes these hydrogen-free compact binaries from their cataclysmic variable cousins is (obviously) the chemical composition and the shorter orbital periods. Both can affect the outburst properties of AM~CVn stars. However, because of the still unknown mechanism driving the in this context necessary variations of viscosity (or the angular-momentum and disc-heating mechanisms), it is difficult to assess the importance of helium richness in the outburst processes. They are potentially important, especially for very low metallicities, but their real impact must await a better understanding of the basic mechanisms driving accretion in discs in general. 

The main conclusion of our study is that outbursts of AM~CVn stars can be explained by the suitably adapted dwarf-nova disc instability model. As in the case of the application of this model to hydrogen-dominated cataclysmic variables, one has to resort to additional effects to account for the observed superoutbursts, dips, cycling states, and standstills. The basic feature that must be included in the model is the variable mass-transfer rate. In particular, we have shown that the enhanced mass-transfer rate, due presumably to variable irradiation of the secondary, must not only be taken into account but is a factor that determines the shape of  most AM~CVn star outburst light-curves. The cause of variable secondary's irradiation has yet to be understood; the best candidate is the precession of a tilted/warped disc. We postulate the existence of super-outbursting Z~Cam type AM CVn stars that have been tentatively identified in observations  \citep[][]{Kato2001}.

The very short periods and low mass-ratios of AM~CVns imply the absence of a pure normal-outburst cycle -- the test bench of the model.
The compactness of AM~CVn stars is clearly important because it accentuates the interactions between the binary components. Their mass ratios $0.0125 \leq q \leq 0.18$ suggest the importance of resonances and tidal effects in general. Irradiation is important as well. Unfortunately, as in the case of cataclysmic variables, the implications of these effects on the outbursts is uncertain and subject to controversy. At best, they can be parameterized only in a fairly rough way. The only effect that, presumably important in CVs, can be neglected in  AM~CVn stars is the irradiation of the inner disc regions by the hot accreting white dwarfs. 

Another drawback that might soon be overcome is the usually unsatisfactory quality of observations of the AM~CV outbursts. When better and richer sets of data will become available the comparison of the DIM with observations will become a precious source of knowledge about accretion disc physics.

\acknowledgements{}
We are grateful to Chris Deloye for providing us with evolutionary tracks for AM~CVn stars and to Joe Patterson for the CR~Boo and V803~Cen
light curves and for helpful comments, to Gavin Ramsay for the light curve of KL Dra and to Eran Ofek and David Levitan for very helpful discussions about the AM~CVn stars observed by the Palomar Transient Factory. We thank the anonymous referee for his report that helped improving the article.
This work has been supported in part by the Polish MNiSW grants PSP/K/PBP/000392 and N N203 380336, the French Space Agency CNES,
and the European Commission via contract ERC-StG-200911.


\begin{appendix}

\section{Critical parameters for various chemical compositions of helium dominated discs.}
\label{app1}

I this appendix we summarize the critical parameter formulae obtained for various chemical compositions of helium-dominated accretion discs.
These formulae have been obtained through fits to numerically calculated S-curves. One should note that whereas the fit of the critical $\Teff$ is accurate within few percent, the
accuracy of the fit to the critical central temperature is no better than 20\%.
\\

\noindent 1. $Y=1$

\begin{eqnarray}
\Sigma^+ & = & 528~\alpha_{0.1}^{-0.81}~R_{10}^{ 1.07}~M_1^{-0.36} \mathrm{g\,cm^{-2}}\nonumber \\
 \Sigma^-  & = &1620~\alpha_{0.1}^{-0.84}~R_{10}^{ 1.19}~M_1^{-0.40}\mathrm{g\,cm^{-2}}\nonumber\\
 T_{\rm c}^+    &=&77000 ~\alpha_{0.1}^{-0.20}~R_{10}^{ 0.08}~M_1^{-0.03}\mathrm{K}\nonumber\\
 T_{\rm c}^-    &=&17800~\alpha_{0.1}^{-0.13}~R_{10}^{-0.03}~M_1^{ 0.01}\mathrm{K}\\
 T_{\rm eff}^+  &=&13000~\alpha_{0.1}^{-0.01}~R_{10}^{-0.08}~M_1^{ 0.03}\mathrm{K}\nonumber\\
 T_{\rm eff}^-  &=&9700~\alpha_{0.1}^{-0.01}~R_{10}^{-0.09}~M_1^{ 0.03}\mathrm{K}\nonumber\\
 \dot{M}^+ &=&1.01\times 10^{17}~\alpha_{0.1}^{-0.05}~R_{10}^{ 2.68}~M_1^{-0.89}\mathrm{g\,s^{-1}}\nonumber\\
 \dot{M}^- &=&3.17\times 10^{16}~\alpha_{0.1}^{-0.02}~R_{10}^{ 2.66}~M_1^{-0.89}\mathrm{g\,s^{-1}}\nonumber 
 \label{hecrit}
 \end{eqnarray}

\noindent 2. Y=0.98, Z=0.02

\begin{eqnarray}
\Sigma^+ & =&380~\alpha_{0.1}^{-0.78}~R_{10}^{ 1.06}~M_1^{-0.35}\mathrm{g\,cm^{-2}}\nonumber \\
\Sigma^-  &=&612~\alpha_{0.1}^{-0.82}~R_{10}^{ 1.10}~M_1^{-0.37}\mathrm{g\,cm^{-2}}\nonumber \\
T_{\rm c}^+    &=&71400~\alpha_{0.1}^{-0.21}~R_{10}^{ 0.08}~M_1^{-0.03}\mathrm{K}\nonumber\\
T_{\rm c}^-    &=& 23600~\alpha_{0.1}^{-0.14}~R_{10}^{-0.00}~M_1^{ 0.00}\mathrm{K}\\
T_{\rm eff}^+  &=&11500~\alpha_{0.1}^{-0.01}~R_{10}^{-0.08}~M_1^{ 0.03}\mathrm{K}\nonumber\\
T_{\rm eff}^-  &=& 8690~\alpha_{0.1}^{-0.00}~R_{10}^{-0.09}~M_1^{ 0.03}\mathrm{K}\nonumber\\
\dot{M}^+ &=& 6.22\times 10^{16}~\alpha_{0.1}^{-0.05}~R_{10}^{ 2.67}~M_1^{-0.89}\mathrm{g\,s^{-1}}\nonumber\\
\dot{M}^- &=& 2.04\times 10^{16}~\alpha_{0.1}^{-0.02}~R_{10}^{ 2.62}~M_1^{-0.87}\mathrm{g\,s^{-1}}\nonumber
\label{98crit}
\end{eqnarray}

\noindent 3. Y=0.96 Z=0.04

\begin{eqnarray}
 \Sigma^+ & =&322~\alpha_{0.1}^{-0.78}~R_{10}^{ 1.04}~M_1^{-0.35}\mathrm{g\,cm^{-2}}\nonumber \\
  \Sigma^-  &=&459~\alpha_{0.1}^{-0.81}~R_{10}^{ 1.08}~M_1^{-0.36}\mathrm{g\,cm^{-2}}\nonumber \\
   T_{\rm c}^+    &=&66800~\alpha_{0.1}^{-0.22}~R_{10}^{ 0.07}~M_1^{-0.02}\mathrm{K}\\
    T_{\rm c}^-   & =&25100~\alpha_{0.1}^{-0.14}~R_{10}^{ 0.00}~M_1^{-0.00}\mathrm{K}\nonumber\\
     T_{\rm eff}^+  &=&10700~\alpha_{0.1}^{-0.01}~R_{10}^{-0.09}~M_1^{ 0.03}\mathrm{K}\nonumber\\
      T_{\rm eff}^-  &=&8350~\alpha_{0.1}^{-0.00}~R_{10}^{-0.10}~M_1^{ 0.03}\mathrm{K}\nonumber\\
       \dot{M}^+ &=&4.76\times 10^{16}~\alpha_{0.1}^{-0.06}~R_{10}^{ 2.65}~M_1^{-0.88}\mathrm{g\,s^{-1}}\nonumber\\
        \dot{M}^- &=&1.74\times 10^{16}~\alpha_{0.1}^{-0.02}~R_{10}^{ 2.61}~M_1^{-0.87}\mathrm{g\,s^{-1}}\nonumber
         \label{96crit} 
 \end{eqnarray}
 
\section{Amplitudes and recurrence times for a selection of U Gem type dwarf novae and outbursting AM CVn stars.}
 
\begin{table}[h]
{\begin{tabular}{|c|c|c|c|}
\hline 
 & $P_{\rm orb}$ [hr] & $A_{n}$ [mag] & $T_{\rm n}$ [d] \tabularnewline
\hline 
IP Peg & $3.79$ & $3.5$ & $95$ \tabularnewline
\hline 
AR And & $3.91$ & $5.9$ & $25$ \tabularnewline
\hline 
UU Aql & $3.92$ & $5.1$ & $48$ \tabularnewline
\hline 
CW Mon & $4.24$ & $3.5$ & $150$ \tabularnewline
\hline 
U Gem & $4.25$ & $5.2$ & $132$ \tabularnewline
\hline 
TW Vir & $4.38$ & $3.2$ & $28$ \tabularnewline
\hline 
SS Aur & $4.39$ & $4.5$ & $51$ \tabularnewline
\hline 
EX Dra & $5.04$ & $3.0$ & $20$ \tabularnewline
\hline 
CZ Ori & $5.25$ & $4.4$ & $26$ \tabularnewline
\hline 
BV Pup & $6.36$ & $2.1$ & $30$ \tabularnewline
\hline 
BF Eri & $6.50$ & $3.0$ & $64$ \tabularnewline
\hline 
SS Cyg & $6.60$ & $3.9$ & $50$ \tabularnewline
\hline 
AF Cam & $7.78$ & $3.6$ & $75$ \tabularnewline
\hline 
MU Cen & $8.21$ & $2.3$ & $45$ \tabularnewline
\hline 
CH UMa & $8.24$ & $4.0$ & $335$ \tabularnewline
\hline 
RU Peg & $8.99$ & $3.7$ & $76$ \tabularnewline
\hline 
AT Ara & $9.01$ & $3.8$ & $97$ \tabularnewline
\hline 
DX And & $10.57$ & $3.9$ & $300$ \tabularnewline
\hline 
EY Cyg & $11.02$ & $4.1$ & $96$ \tabularnewline
\hline 
V442 Cen & $11.04$ & $3.5$ & $32$ \tabularnewline
\hline 
BV Cen & $14.67$ & $1.9$ & $167$ \tabularnewline
\hline 
\hline
CR Boo & $0.41$ & $1.2$ & $1$ \tabularnewline
\hline 
V803 Cen & $0.44$ & $1.3$ & $0.95$ \tabularnewline
\hline 
PTF1J0719 & $0.45$ & $2.5$ & $10$ \tabularnewline
\hline 
\end{tabular}}
\par
\centering{}
\caption{\footnotesize The list of U Gem-type and AM CVn systems (three last rows) plotted on Fig. \ref{fig:KPobs}. The columns are: $P_{\rm orb}$ is the orbital period, $A_{n}$ is the normal outburst amplitude and $T_{\rm n}$ is the normal outburst reccurence time. Data from \citet{Warner03} and \citet{Ritter03}.}
\label{tab:UGem_KP}
\end{table}
  
\end{appendix}

\end{document}